\documentclass[11pt]{article}
\usepackage{epsfig,amsfonts,amssymb}
\usepackage{amsmath,graphics}
\usepackage{slashed}
\usepackage{float}
\usepackage[utf8]{inputenc}
\usepackage{scalerel}
\usepackage{graphicx}
\usepackage{stackengine}
\usepackage{hyperref}
\usepackage{cleveref}
\usepackage{amsmath}
\usepackage{amsmath,mleftright}
\usepackage{xparse}
\usepackage{slashed}
\usepackage{amssymb}
\usepackage{cancel}
\usepackage{multirow}
\usepackage[utf8]{inputenc}
\usepackage{color}   
\usepackage{braket}
\usepackage[vcentermath]{youngtab}
\usepackage{tikz}
\usetikzlibrary{decorations.pathreplacing}
                                                   
\usepackage{xcolor}
\usepackage{bbm}
\usepackage{amsmath}
\usepackage{url}
\usepackage{longtable}
\usepackage{array}
\usepackage{bbold}
\numberwithin{equation}{section}

\newcommand{\ads}{\mathrm{AdS_4}}

\newcommand{\half}{\frac{1}{2}}

\newcommand{\Tr}{\mathop{\mathrm{tr}}}

\newcommand{\g}{\dagger}
\newcommand{\NN}{\mathcal{N}}
\newcommand{\WW}{\mathcal{W}}

\newcommand{\ZZ}{\mathcal{Z}}

\newcommand{\nn}{\nonumber}

\newcommand{\ben}{\begin{eqnarray*}}
	\newcommand{\en}{\end{eqnarray*}}

\newcommand{\superN}{\mathcal{N}}

\ifx\href\asklfhas\newcommand{\href}[2]{#2}\fi
\ifx\arxivref\asklfhas\newcommand{\arxivref}[1]{\href{http://arxiv.org/abs/#1}%
	{#1}}\fi
\ifx\doiref\asklfhas\newcommand{\doiref}[2]{\href{http://dx.doi.org/#1}{#2}}\fi

\newcommand{\be}{\begin{eqnarray}}
	\newcommand{\ee}{\end{eqnarray}}

\usepackage{url}
\usepackage{longtable}

\usepackage{hyperref}
\def\one{{\hbox{ 1\kern-.8mm l}}}
\def\zero{{\hbox{ 0\kern-1.5mm 0}}}

\begin{document}
	
	\baselineskip 24pt
	
	\begin{center}
		{\Large \bf Mass spectrum in $\ads\times U(1)\setminus U(3)/U(1) $ compactification using harmonic analysis}

	\end{center}
	
	\vskip .6cm
	\medskip
	
	\vspace*{4.0ex}
	
	\baselineskip=18pt

	\centerline{\large \rm   Moumita Patra}
	
	\vspace*{4.0ex}
	
	\centerline{\large \it  Department of Physics, Indian Institute of Technology,Kharagpur,}
	
	\centerline{\large \it  Kharagpur, 721302,  India}
	\vspace*{1.0ex}
	
	%\centerline{\large \it Homi Bhabha National Institute, Training School Complex,}
	
	%\centerline{\large \it  Anushakti Nagar, Mumbai, India 400085}
	
	\vspace*{4.0ex}
	\centerline{E-mail:  mpatra91@gmail.com }
	
	\vspace*{5.0ex}

	\centerline{\bf Abstract} \bigskip
	\thispagestyle{empty}
	In this paper we present the Kaluza-Klein mass spectrum of  0-form and 1-form  fields that appear in $\ads\times B_7$ compactification of $d=11$ supergravity, where, $B_7=U(1)\setminus U(3)/ U(1)$. This theory belongs to the  class of gravity duals to  three dimensional quiver Chern-Simons matter theories as demonstrated by D.L Jafferis and A. Tomasiello in \cite{jaff}.

	\vfill \eject
	
	\baselineskip=18pt
	
	\newpage
	\setcounter{page}{1}
	\renewcommand{\thefootnote}{\arabic{footnote}}
	\setcounter{footnote}{0}
	
	\hrule
	\tableofcontents
	\vspace{8mm}
	\hrule
	\vspace{4mm}

	%\section{corrections}

	\section{\label{1}Introduction}
	Kaluza-Klein(KK) compactification \cite{Kaluzha-klein} of $d=11$ supergravity attracted significant interest  in the 1980's due the following features.\\
	a) In $d=11$ the maximal supersymmetry one can obtain is  $\NN=1$, which makes the theory simple and compact. \\
	b) Gravity theory on 11 dimensional  manifold $M_{11}$ splits into $M_4\otimes M_7$ \cite{Witten}, where $M_4$ is anti-de Sitter space and $M_7$ is any 7 dimensional Einstein space.  \\
	c) The effective theory on $M_4$ contains graviton, gravitino, photon that appear to naturally interpret as gauge fields capturing the geometry of $d=11$ superspace.\\	In this paper we present the KK mass spectrum of 0-form and 1-form fields that appear in $\ads\times M_7$ compactification of $d=11$ supergravity, specifying the Einstein manifold $M_7$. 
	\par From a modern perspective, the primary motivation for determining the supergravity mass spectrum is rooted in the AdS/CFT duality. Matching the superconformal index has been a well known and successful  method to verify gravity dual of several superconformal field theories \cite{Kim:2009wb}\cite{Bhattacharya:2008bja}. The key ingredient in computing the superconformal index on the gravity side is the graviton mass spectrum, which corresponds to representations of the associated superconformal algebra. A commonly used approach to derive the graviton mass spectrum is the harmonic analysis technique.
	However, while extensive research exists for cases involving homogeneous $ M_7$ \cite{round7a}-\cite{Termonia:1999cs}, considerably less literature is available, to our knowledge, when $ M_7 $ is a non-homogeneous Einstein manifold.
	In this note we attempt to extend the technique of harmonic analysis involving non-homogeneous Einstein manifolds.
	 
\par Now, let us provide the motivation for this work framed within the context of a review of the relevant literature.
Three dimensional Chern Simons matter theories are known to have M-theory dual in the form $\ads\times M_7$, where $M_7$ is some Sasaki-Einstein manifold. The renowned  example is the ABJM\cite{abjm} theory whose gravity dual  is $\ads\times S^7/Z_k$ where $k$ is the Chern-Simons level of the theory. This conjecture has undergone various tests and has been well established in the literature\cite{Kim:2009wb}.
$\NN=3$ quiver Chern-Simons theory,  examined in this paper reduces to the ABJM theory when restricted to two gauge groups. In other words ABJM theory is a special case of $\NN=3$ circular quiver Chern-Simons theory.
 Restricting to the case with two gauge groups enhances the supersymmetry of the theory making it $\NN=6$, which can be further enhanced to $\NN=8$ if one considers $k=1,2$. $\NN=8$ is the maximal supersymmetry one can obtain in three dimensions.
The bigger supersymmetry is sometimes helpful in  computing various quantities in a theory. To elaborate,
d=11 supergravity compactified on $\ads\times M_7$ where, $M_7$ has $\NN$ supersymmetry, the KK spectrum of the operators are the unitary irreducible representations(UIR) of the algebra $OSp(\NN|4)\times G^\prime$. $G^\prime$ is factor in the isometry group of the associated 7-manifold that commutes with supersymmetry and with the R-symmetry factor $SO(\NN)$. Here, $OSp(\NN|4)$ is the the global symmetry group of $\ads$. The bosonic part of $OSp(\NN|4)$ is $Sp(4,\mathbbm{R})\equiv SO(3,2)$ and the fermionic part is $SO(\NN)$. Also, the isometry group $G$ of $M_7$ can be expressed as,
\be 
G=SO(\NN)\times G^\prime .
\ee 
In case of $\NN=8$ supersymmetry $G=SO(\NN)$ which coincides with the R-symmetry group of the superconformal field theory. Therefore $G^\prime $ is absent in the  case with $\NN=8$. This fact makes it possible the derive the KK spectrum only by studying UIR of $OSp(8|4)$\cite{Bhattacharya:2008bja}\cite{Gunaydin:1985tc}.

\par  However, in cases with lower supersymmetry,  the Kaluza-Klein (KK) states must be organized into supermultiplets of $OSp(\NN|4)\times G^\prime$.
For our case with $\NN=3$, we have  $G=U(2)\times U(1)$. 
 Consequently, the KK states must be organized into supermultiplets of $OSp(3|4) \times U(1) \times U(1)$. In this scenario, the KK states are not unitary irreducible representations (UIRs) of the $OSp(3|4)$ superalgebra alone. 
Therefore in order to calculate the mass spectra of gravitons on $B_7$ we need to study the unitary irreducible representations of the $OSp(3|4)\times U(1)\times U(1)$ algebra.

\par This work is part of the effort to match the superconformal index computed in \cite{Patra:2021dbu} for $\NN=3$ quiver Chern-Simons theory with $\widehat{ADE}$ classification. 
In \cite{Patra:2021dbu}, we computed the superconformal index of the conformal field theory, leaving the computation of the same in  conjectured dual gravity theory on $\ads\times B_7$\cite{jaff}. To achieve this, one must determine the spectrum of fields in the M-theory dual, which is obtained through harmonic analysis. Below we list the key features of the harmonic analysis technique.
	\begin{enumerate}
		\item \textbf{Identify the Compactification Manifold}: Since \(d=11\) supergravity typically involves compactification involving an internal manifold(e.g., $S^7$ or $AdS_4 \times S^7$), first specify the compact manifold on which the extra dimensions are compactified. This will affect the spectrum and symmetry of the theory.
		
		\item \textbf{Decompose Fields into Harmonics}: Decompose the \(d=11\) fields on the compactification manifold using its harmonic functions. For example, if compactifying on $S^7$, use spherical harmonics. This step breaks down each field into modes with distinct masses.
		
		\item \textbf{Apply Eigenvalue Equations}: Use the Laplacian or other differential operators on the compact manifold to derive eigenvalue equations for each harmonic component, as each eigenvalue corresponds to a possible mass in lower-dimensional spacetime.
		
		\item \textbf{Determine the Mass Spectrum}: Solve the eigenvalue equations to obtain the discrete set of eigenvalues. These eigenvalues correspond to the mass spectrum of the compactified theory.
		
		\item \textbf{Organize in Supermultiplets}: Arrange the resulting mass states according to the supersymmetry of the lower-dimensional theory, identifying them with specific supermultiplets.
	\end{enumerate}
	This approach enables us to derive the mass spectra of fields after compactification in a way that aligns with 
	AdS/CFT duality principles, crucial for matching states across the duality. 
	
 The rest of the paper is organized as follows. In section \ref{sec:diff_geo} we study various aspects  of differential geometry of $B_7$ relevant to our computation. In \ref{sec:harmonic_analysis} we construct the harmonics and study relevant representations in order to obtain the mass spectrum. We summarize our results and provide future directions in \ref{sec:conclusion}. Appendix \ref{appendix:A} contains all the necessary conventions used in the calculations. Lastly, we present the computation of Ricci tensor in \ref{appendix:B}.
\section{Studying the manifold $B_7$}\label{sec:diff_geo}
	We start this section by  briefly reviewing the manifold $B_7$, obtained in \cite{jaff}, as a gravity dual to $\NN=3$ $\widehat{A}_2$  CS theory. This is a quiver Chern-Simons matter theory with gauge group $U(N)_{k_1}\times U(N)_{k_2}\times U(N)_{k_3}$, with vanishing sum of CS levels $\sum_{i=1}^3 k_i=0$.
	 The moduli space of such a theory, denoted by $\mathcal{M}_{n=3}$,
	 \be
	 \label{moduli_space}
	 \mathcal{M}_{n=3}=\frac{\{q_j=(\WW_j,\bar{\ZZ}_j)^T\in \mathbbm{C}^{2n}|\sum_{j}M_{ji}q^\g_{i}\sigma_\alpha q_{i}=0  \} }{(\WW_j, \ZZ_j)\sim (\mathbf{n}_j \WW_j, \mathbf{n}_j^{-1} \ZZ_j), \mathbf{n}\in \mathbf{N}}\; \qquad i,j=1,2,3
	 \ee
	 is an eight dimensional toric hyper-K\"{a}hler manifold which is a cone over $B_7$. $({\ZZ}_j,\WW_j)$ are the complex scalars of the hypermultiplet, in the bi-fundamental representation under $(U(N)_i, U(N)_{i+1} ) $. The geometry of $\mathcal{M}_{n=3}$ is studied via hyper-K\"{a}hler quotient construction and is shown that $\mathcal{M}_{n=3}$ is a cone over 
	 \be 
	 B_7\equiv U(1)_L\setminus U(3)/ U(1)_R
	 \ee 
	 The action of $U(1)_L$ is given by, $\mathrm{diag}(e^{ik^\prime_1\theta},e^{ik^\prime_2\theta},e^{ik^\prime_3\theta})$, where $k_1^\prime=k_1,k_2^\prime=k_2,k_3^\prime=-k_3$ and the action of $U(1)_R$ is given by $\mathrm{diag}(1,1,e^{i\varphi})$. Note that neither of the $U(1)$'s has determinant 1, since $k^\prime_1+k^\prime_2+k^\prime_3\neq 0$. Such manifolds are called double coset manifolds and it can be shown that there are 7 distinct equivalence classes defined by,
	 \be 
	 \{U(1)_L\; g\; U(1)_R\;|\;g\in U(3)\}\; .
	 \ee 
Such manifolds are referred to as bi-quotients with strictly positive curvature and have been  studied in the literature\cite{BGM1}\cite{Eschenburg}.
	\subsection{Differential geometry on $B_7$}
	Now, to carry out the computations we work with the algebra elements of $B_7$. Therefore let us choose a basis of $b_7$ and note all the commutation relations among them. First, we write $u(3)\cong su(3)\times u(1)$ whose generators are the standard Gell-Mann matrices with the  addition of an $u(1)$ element.
	The generators of 9 dimensional $u(3)$ algebra are,
	\be 
	\{\frac{i}{2}\lambda_1,\frac{i}{2}\lambda_2,\frac{i}{2}\lambda_3,\frac{i}{2}\lambda_4,\frac{i}{2}\lambda_5,\frac{i}{2}\lambda_6,\frac{i}{2}\lambda_7,\frac{i}{2}\lambda_8,\frac{i}{2}\lambda_9 \}.
	\ee 
whose explicit matrix representations are given in \ref{appendix:A}.
	The $U(1)_L$ group element is
	\be 
	U(1)_L&=&	\begin{pmatrix}
		e^{ik_1^\prime\theta }& 0& 0\\0&	e^{ik_2^\prime\theta }& 0\\	0&0&e^{ik^\prime_3\theta }
	\end{pmatrix},\quad k^\prime_1,k^\prime_2,k^\prime_3\in \mathbbm{R},k^\prime_1,k^\prime_2,k^\prime_3>0,\det(U(1)_L)\neq 1\nn\\
	U(1)_R&=& \begin{pmatrix}
		1&0&0\\0&1&0\\0&0&e^{i\varphi}
	\end{pmatrix}, \quad  \det(U(1)_R)\neq 1
	\ee 
	Since we carry out the computations in terms of  algebra elements, the above group elements are  converted to the algebra elements given below,
	\be 
	u(1)_L=	\begin{pmatrix}
		{k_1^\prime }& 0& 0\\0&	{k_2^\prime }& 0\\	0&0&{k^\prime_3 }
	\end{pmatrix}\qquad 	u(1)_R= \begin{pmatrix}
	0&0&0\\0&0&0\\0&0&1
	\end{pmatrix} \qquad k_1^\prime+k_2^\prime-k_3^\prime=0
	\ee 
	obtained by differentiating the group element, $g=e^{ih\theta},ih=\frac{dg}{d\theta}|_{\theta=0}$. 
	\par Now, to define the embedding of $u(1)_L$ and $ u(1)_R$ in $u(3)$ we take a linear combination of the Cartans of $u(3)$ as follows,
	\be 
	Z=k_1^\prime \frac{i}{2}\lambda_3+k_2^\prime\frac{i}{2}\sqrt{3}\lambda_8+k_3^\prime\sqrt{\frac{{3}}{{2}}}\frac{i}{2}\lambda_9
	\ee 
	Note that we feed the data which is given by the set of three positive integers $(k^\prime_1,k^\prime_2,k^\prime_3)$ of the moduli space that comes from the CFT side, into the gravity side as above. Next,  we impose the following constraints,
\be 
\Tr(ZZ_L)=\Tr(ZZ_R)=\Tr(Z_LZ_R)=0\quad Z_L\in u(1)_L, Z_R\in u(1)_R
\ee 
The above orthogonality conditions ensures that the following decomposition of the $u(3)$ algebra holds,
\be 
u(3)=u(1)_L\oplus u(1)_R\oplus b_7
\ee 
This decomposition also implies that invariant one forms on $U(3)$
\be 
\Omega=g^{-1}dg\qquad g\in u(3)
\ee 
also can be decomposed as,
\be 
\Omega=\Omega_L\oplus\Omega_R\oplus\Omega_{b_7}\;.
\ee 
The left invariant one forms  satisfy Maurer-Cartan equations,
\be 
d\Omega+\Omega\wedge \Omega=0
\ee 
Therefore,
\be 
d\Omega=d\Omega_L\oplus d\Omega_R\oplus d\Omega_{b_7}.
\ee 
\be 
&&d\Omega+\Omega\wedge\Omega=d\Omega_L+d\Omega_R+d\Omega_{b_7}+\Big[\Omega_L+\Omega_R+\Omega_{b_7}\Big]\wedge\Big[\Omega_L+\Omega_R+\Omega_{b_7}\Big]=0\nn\\
&&\implies d\Omega_L+d\Omega_R+d\Omega_{b_7}+\Big[\Omega_L+\Omega_R+\Omega_{b_7}\Big]\wedge\Big[\Omega_L+\Omega_R+\Omega_{b_7}\Big]=0\nn\\
&&\hspace*{-1cm}\implies d\Omega_L+d\Omega_R+d\Omega_{b_7}+\Omega_L\wedge \Omega_R+\Omega_L\wedge \Omega_{b_7}+\Omega_R\wedge\Omega_L+\Omega_R\wedge\Omega_{b_7}+\Omega_{b_7}\wedge \Omega_{L}+\Omega_{b_7}\wedge \Omega_{R}+\Omega_{b_7}\wedge\Omega_{b_7}=0\nn\\
&&\hspace*{-1cm}\implies d\Omega_{b_7}+\Omega_{b_7}\wedge\Omega_{b_7}=0
\ee 
since,
\be 
d\Omega_L=0\qquad d\Omega_R=0
\ee 

Now, let us take
$k_1^\prime=1, k_2^\prime=2, k_3^\prime=3$, which gives,
	\be 
Z &=&  \frac{i}{2}\; \lambda_3 + i \sqrt{3}\; \lambda_8 + \frac{3i \sqrt{3} }{2\sqrt{2}}\;
\lambda_9\nn\\
Z_L &=&  \frac{i}{2}\; \lambda_3 +  i \sqrt{3}\; \lambda_8 - \frac{13i  \sqrt{3} }{9 \sqrt{2}}\; \lambda_9	\nn\\
Z_R &=& -6 i\; \lambda_3 + i \sqrt{3}\; \lambda_8 \;.
\ee
The generators of $b_7=h_L\setminus g/ h_R$ algebra are(where, $g=u(3)$),
\be 
\label{eqn:b7_generator}
\{\frac{i}{2}\lambda_1,\frac{i}{2}\lambda_2,\frac{i}{2}\lambda_4,\frac{i}{2}\lambda_5,\frac{i}{2}\lambda_6,\frac{i}{2}\lambda_7, Z\}
\ee
which makes it a 7-dimensional algebra.
 $B_7$ being a Lie group  can be parametrized by defining a co-ordinate system $\{y_\alpha\},\alpha=1,2,4,5,6,7,Z$, hence becomes a  group manifold. The coset representative,
\be 
L(y_\Lambda)=e^{\sum_{\Lambda=1}^9 i\lambda _\Lambda y^\Lambda}
\ee 	
is used to construct a left invariant one form on $u(3)$ as follows,
\be 
\Omega(y)=L^{-1}(y)\, dL(y)=\Omega^\Lambda(y)\, T_\Lambda,\quad \Lambda\in u(3),\Lambda=1,...9
\ee 
$\Omega(y)$ satisfies the Maurer-Cartan equation, which follows from the identity $d(L^{-1}L)=0$,
\be 
\label{eqn:maurer_cartan}
d\Omega^\Lambda+\frac{1}{2}\mathcal{C}^\Lambda_{\Pi\Sigma}\, \Omega^\Pi\wedge \Omega^\Sigma=0
\ee 
Note that $\mathcal{C}^\Lambda_{\Pi\Sigma}$ is the structure constant on $SO(7)$ whose indices are raised and lowered by the $SO(7)$ metric $\eta^{\alpha\beta}=diag(-,-,-,-,-,-,-)$. On the other hand $f^{\Lambda\Pi\Sigma}$ are the structure constants on flat space. Upon writing \eqref{eqn:maurer_cartan}  explicitly one finds  the following Maurer-Cartan equations
\be 
&&\boxed{d\Omega^m-f_{mn3}\, \Omega^n\wedge \Omega^3-f_{mA\dot{B}}\, \Omega^A\wedge \Omega^{\dot{B}}=0}\qquad m,n=1,2\nn\\
&&\boxed{d\Omega^3-\frac{1}{2} f_{3mn}\, \Omega^m\wedge \Omega^n-\frac{1}{2}f_{3AB}\, \Omega^A\wedge \Omega^B-\frac{1}{2}f_{3\dot{A}\dot{B}}\, \Omega^{\dot{A}}\wedge \Omega^{\dot{B}}=0}\nn\\
&&\boxed{d\Omega^A+f^{mA\dot{B}}\,\Omega_{m}\wedge \Omega_{\dot{B}}+f^{3AB}\,\Omega_{3}\wedge \Omega_B+f^{8AB}\,\Omega_{8}\wedge \Omega_B=0}\qquad A=4,5\nn\\
&&\boxed{d\Omega^{\dot{A}}+f^{m\dot{A}{B}}\,\Omega_{m}\wedge \Omega_{{B}}+f^{3\dot{A}\dot{B}}\,\Omega_{3}\wedge \Omega_{\dot{B}}+f^{8\dot{A}\dot{B}}\,\Omega_{8}\wedge \Omega_{\dot{B}}=0}\qquad \dot{A},\dot{B}=6,7\nn\\
&&\boxed{d\Omega^8-\half f^{8AB}\Omega_{A}\wedge \Omega_B-\half f^{8\dot{A}\dot{B}}\Omega_{\dot{A}}\wedge \Omega_{\dot{B}}=0}\nn\\
&&\boxed{d\Omega^9=0}\nn\\
&&\boxed{d\Omega^Z-\frac{1}{2} f_{3mn}\, \Omega^m\wedge \Omega^n-\frac{1}{2}f_{3AB}\, \Omega^A\wedge \Omega^B-\frac{1}{2}f_{3\dot{A}\dot{B}}\Omega^{\dot{A}}\wedge \Omega^{\dot{B}}
	-\sqrt{3}f^{8AB}\Omega_{A}\wedge \Omega_B-\sqrt{3} f^{8\dot{A}\dot{B}}\Omega_{\dot{A}}\wedge \Omega_{\dot{B}}=0}\;.\nn\\
\ee  
Since we want to find a manifold that is Einstein, we rescale the 
 the vielbeins on $b_7$ as follows,
\be 
\mathcal{B}^m&=&\frac{1}{a}\,\Omega^m\qquad m=1,2\nn\\
\mathcal{B}^A&=&\frac{1}{b}\,\Omega^A\qquad A=4,5\nn \\
\mathcal{B}^{\dot{A}}&=&\frac{1}{\dot{b}}\,\Omega^{\dot{A}}\qquad \dot{A}=6,7\nn \\
\mathcal{B}^Z&=&\frac{1}{c}\,\Omega^Z
\ee 
Now the task is to solve the Einstein equation for $a,b,\dot{b},c$. 
The above vielbeins obey the following two properties.
\begin{itemize}
	\item \underline{Invariance under the isometry group of $B_7$:}
	$\mathcal{B}^\alpha$ are invariant under $U(2)\times U(1)$, i.e
	\be 
	l_\Lambda \mathcal{B}^\alpha:=l_{k_\Lambda}\mathcal{B}^\alpha=W_\Lambda^{\alpha\beta}\mathcal{B}_\beta
	\ee 
	where,
	$k_\Lambda$ is a tangent vector on $b_7$, $l_\Lambda$ is the Lie derivative along $k_\Lambda$. $W_\Lambda^{\alpha\beta}=-W_\Lambda^{\beta\alpha}\in so(7)$.
	
	\item \underline{Einstein}: We impose that $\mathcal{B}^\alpha$ correspond to an Einstein metric on $b_7$. To achieve that 
	we impose torsion free condition on $b_7$ through the following equation,
	\be 
	\label{eqn:no_torsion}
	d \mathcal{B}^\alpha-\mathcal{B}^{\alpha\beta}\wedge \mathcal{B}_\beta=0\;.
	\ee
	After that we compute Riemann curvature tensor is defined as,
	\be 
	\mathcal{R}^{\alpha\beta}=d\mathcal{B}^{\alpha\beta}-\mathcal{B}^{\alpha\gamma}\wedge \mathcal{B}^\beta_\gamma=d\mathcal{B}^{\alpha\beta}-\eta_{\gamma\sigma}\,\mathcal{B}^{\alpha\gamma}\wedge \mathcal{B}^{\sigma\beta}:=\half  \mathcal{R}^{\alpha\beta}_{\gamma\delta}\, \mathcal{B}^\gamma\wedge \mathcal{B}^\delta
	\ee 
	and solve for \cite{noi321}\cite{Boyer:1998sf}
	\be 
	\mathcal{R}^{\alpha}_{\beta}=12\delta^\alpha_\beta,
	\ee 
 where $\mathcal{R}^{\alpha\beta}_{\gamma\delta}$ is the Riemann tensor.
	
\end{itemize}	
The first step for evaluating the Ricci tensor $\mathcal{R}^{\alpha}_{\beta}$ involves determining the antisymmetric spin connections $\mathcal{B}^{\alpha\beta}$. We present the results below, while the details of the computation is given in appendix \ref{appendix:B}.
\be \label{eqn:spin_connection}
&&\mathcal{B}^{AB}=-f^{3AB}\,\Omega_{3}-f^{8AB}\Omega_{8}-(\frac{b^2}{2c}f_{3AB}+\frac{\sqrt{3}b^2}{c}f^{8AB})\mathcal{B}^Z \nn\\
&&\mathcal{B}^{A\dot{B}}=\Big(-\frac{b\dot{b}}{2a}f^{mA\dot{B}}
+\frac{a\dot{b}}{2b} f^{mA\dot{B}}
+\frac{ab}{2\dot{b}}f^{mA\dot{B}}\Big)\mathcal{B}^m\nn\\
&&\mathcal{B}^{\dot{A}\dot{B}}=- f^{3\dot{A}\dot{B}}\,\Omega_{3}-f^{8\dot{A}\dot{B}}\,\Omega_{8}
-(\frac{\dot{b}^2}{2c}f^{3\dot{A}\dot{B}}
+\frac{\sqrt{3}\dot{b}^2}{c} f^{8\dot{A}\dot{B}}) \mathcal{B}^Z\nn\\
&&\mathcal{B}^{mA}=
\Big[\frac{b\dot{b}}{2a}f^{mA\dot{C}}
+\frac{a\dot{b}}{2b} f^{mA\dot{C}}
+\frac{ab}{2\dot{b}}f^{m\dot{C}{A}}\Big]\mathcal{B}^{\dot{C}}\nn\\
&&\mathcal{B}^{m\dot{B}}=\Big[-\frac{b\dot{b}}{2a}f^{mA\dot{B}}
-\frac{a\dot{b}}{2b} f^{m\dot{B}A}
+\frac{ab}{2\dot{b}}f^{m\dot{B}{A}}\Big]\mathcal{B}^A\nn\\
&&\mathcal{B}^{AZ}=-(\frac{b^2}{2c}f^{3AB}+\frac{\sqrt{3}b^2}{c}f^{8AB})\mathcal{B}^{B}\nn\\ 
&&\mathcal{B}^{\dot{A}Z}=-(\frac{\dot{b}^2}{2c}f^{3\dot{A}\dot{B}}
+\frac{\sqrt{3}\dot{b}^2}{c} f^{8\dot{A}\dot{B}})\mathcal{B}^{\dot{B}}\nn\\
&&\mathcal{B}^{mn}=f^{mn3} \Omega^3+\frac{a^2}{2c}f^{3nm} \mathcal{B}^Z\nn\\
&&
\mathcal{B}^{mZ}=\frac{a^2}{2c}f^{3nm} \mathcal{B}^n
\ee 

\subsection{Embedding $U(1)_L, U(1)_R$ in $SO(7)$}
To compute the $SO(7)$ covariant derivative on various differential forms, it is necessary to understand the embedding of $u(1)_L,u(1)_R$ within \(SO(7)\).  The embedding for $SO(7)$ vector representations is given as \cite{fermionauria},
\be 
\label{eqn:embedding}
(T_H)^\alpha_{\;\beta}=\mathcal{C}^\alpha_{H\beta}
\ee
Therefore, for the generators of $(u(1)_L, u(1)_R)$ we find,
\be 
(T_{Z_L})^\alpha_{\;\beta} &=&   -12 \; \mathcal{C}^\alpha_{3\beta} +  4\sqrt{3}\;\mathcal{C}^\alpha_{8\beta}\;   \nn\\
(T_{Z_R})^\alpha_{\;\beta} &=&  \mathcal{C}^\alpha_{3\beta} +  2 \sqrt{3} \;\mathcal{C}^\alpha_{8\beta} - \frac{26  \sqrt{3} }{9 \sqrt{2}}\; \mathcal{C}^\alpha_{9\beta}
\ee 
The structure constant on $U(3)$ can be expanded as,
\be 
\mathcal{C}^\alpha_{Z\beta} &=& \mathcal{C}^\alpha_{3\beta} + 2 \sqrt{3}\; \mathcal{C}^\alpha_{8\beta} + \frac{3 \sqrt{3} }{\sqrt{2}}\;\cancelto{0}{\mathcal{C}^\alpha_{9\beta}}
\ee 
Note that, for $\alpha=\beta$ the entries of $(T_H)^\alpha_\beta$ will be zero since the structure constants $\mathcal{C}^\alpha_{H\beta}$ are completely antisymmetric. Moreover, the LHS of  equation \eqref{eqn:embedding}  we expand the $\empty^\alpha_{\;\beta}$ as, $(11,12,13,14,15,16,17)$ and so on while on the RHS we actually mean $(11,12,14,15,16,17,1Z)$.
This is the index structure of $\alpha,\beta$ in $(T_H)^\alpha_\beta$, which we evaluate below.	
\be 
(T_{Z_L})^\alpha_{\;\beta}=\begin{pmatrix}
	0&-12&0&0&0&0&0\\
	12&0&0&0&0&0&0\\
	0&0&0&0&0&0&0\\
	0&0&0&0&0&0&0\\
	0&0&0&0&0&12&0\\
	0&0&0&0&-12&0&0\\
	0&0&0&0&0&0&0
\end{pmatrix}
\ee 

\be 
(T_{Z_R})^\alpha_{\;\beta}=\begin{pmatrix}
	0&1&0&0&0&0&0\\
	-1&0&0&0&0&0&0\\
	0&0&0&\frac{7}{2}&0&0&0\\
	0&0&-\frac{7}{2}&0&0&0&0\\
	0&0&0&0&0&\frac{5}{2}&0\\
	0&0&0&0&-\frac{5}{2}&0&0\\
	0&0&0&0&0&0&0
\end{pmatrix}
\ee
For $SO(7)$ spinor representations we use the following embedding of the generators of $u(1)_L,u(1)_R$,
\be 
\label{eqn:spinor_embedding}
T_{Z_L}^{(s)}=-\frac{1}{4}(T_{Z_L})^{\alpha\beta}\;\tau_{\alpha\beta}\qquad T_{Z_R}^{(s)}=-\frac{1}{4}(T_{Z_R})^{\alpha\beta}\;\tau_{\alpha\beta}
\ee 
where, $\tau_{\alpha\beta}=[\tau_\alpha,\tau_\beta]$, $\tau_\alpha$'s are Clifford algebra elements of $SO(7)$, satisfying
\be 
\{\tau_\alpha,\tau_\beta\}=2\eta_{\alpha\beta}=-2\delta_{\alpha\beta}.
\ee 
Explicit matrix forms of $\tau_{\alpha}$'s are given in appendix
\ref{app:A} . Evaluating \eqref{eqn:spinor_embedding} one finds,
\be 
T_{Z_L}^{(s)}=\begin{pmatrix}
	0&0&0&0&0&0&0&0\\
	0&-12i&0&0&0&0&0&0\\
	0&0&0&0&0&0&0&0\\
	0&0&0&-12i&0&0&0&0\\
	0&0&0&0&12i&0&0&0\\
	0&0&0&0&0&0&0&0\\
	0&0&0&0&0&0&12i&0\\
	0&0&0&0&0&0&0&0
\end{pmatrix}\nn\\ [3mm]
T_{Z_R}^{(s)}=\begin{pmatrix}
	0&0&0&0&0&0&0&0\\
	0&	-\frac{5i}{2}&0&0&0&0&0&0\\
	0&0&-\frac{7i}{2}&0&0&0&0&0\\
	0&0&0&i&0&0&0&0\\
	0&0&0&0&\frac{5i}{2}&0&0&0\\
	0&0&0&0&0&0&0&0\\
	0&0&0&0&0&0&-i&0\\
	0&0&0&0&0&0&0&\frac{7i}{2}\\
\end{pmatrix}.
\ee 
At this stage we have gathered  all the necessary information about $b_7$ to carry out the harmonic analysis. 
\section{Harmonic analysis on $b_7$}\label{sec:harmonic_analysis}
Harmonic expansion is a mathematical tool used to analyze fields on compact spaces, particularly in the context of KK compactifications of higher-dimensional supergravity theories. 
The goal of harmonic expansion is to decompose the fields on $b_7$ into eigenfunctions of the  Laplace-Beltrami operator associated with the manifold. 
In this section we introduce the harmonics and discuss the uses of harmonic analysis technique in solving differential equations. The liearized equation of 11d supergravity that we intend to solve is
\be
\label{eqn:sugra_eqn} 
\Big(\Box_x^{[J_1,J_2]} +\boxtimes_y^{[\lambda_1, \lambda_2, \lambda_3]}\Big)\Phi^{[J_1,J_2]}_{[\lambda_1, \lambda_2, \lambda_3]}(x,y)=0\;,
\ee  
where, $\Phi^{[J_1,J_2]}_{[\lambda_1, \lambda_2, \lambda_3]}(x,y)$ is a field on $\ads\times b_7$ transforming in the irrep $[J_1,J_2]$ of $SO(3,2)$ and $[\lambda_1,\lambda_2,\lambda_3]$ of $SO(7)$.
The differential  operator $\boxtimes_y$ on $b_7$ is an invariant operator since it satisfies the following property,
\be 
[\mathcal{L}_\Lambda,\boxtimes_y]=0\qquad \forall \Lambda\in G
\ee 
where, $\mathcal{L}_\Lambda$ is the covariant Lie derivative on $b_7$.

\par Now let us see how to use the harmonic expansion of the fields over $b_7$ solve the above differential equation.
 Any function $\phi(g)$ on a group manifold $G$ can be expanded as,
\be 
\phi(g)=\sum_\mu \sum_{p,q} C_{pq}^{(\mu)}\; D_{pq}^{(\mu)}(g)
\ee 
where, $D_{pq}^{(\mu)}(g)$ are matrix elements of UIR of $G$  labeled by $\mu$.  $D_{pq}^{(\mu)}(g)$'s form complete orthonormal basis on the group manifold $G$. Moving on to a bicoset manifold such as $b_7$, parametrized by $y$, a function $\phi(y)$ on $b_7$ can still be expanded as
\be 
\phi^\beta(L(y))_i=\sum_\mu \sum_{p} C_{p}^{(\mu)}\; D_{pi}^{(\mu)}(L(y))
\ee 
where, $\beta$ indicates that $\phi^\beta(L(y))$ is also an irrep of $U(1)_L\times U(1)_R$ labeled by $\beta$, $i$ is the index in the representation $\beta$. 
\par Now, let us examine  the harmonic expansion of a generic $SO(7)$ irreducible field $\Phi^{[\lambda_1\lambda_2\lambda_3]}(z)$, where $z$ are the co-ordinates of $SO(7)$. An $SO(7)$ representation is labelled by three integer or half-integer numbers, $\lambda_1\geq \lambda_2\geq \lambda_3$. %$\Phi^{[\lambda_1\lambda_2\lambda_3]}(z)$ also contains another index $n=dim[\lambda_1\lambda_2\lambda_3]$. $n$ is defined over $b_7$. 
We denote the complete set of orthonormal basis for the expansion of $\Phi^{[\lambda_1\lambda_2\lambda_3]}(y)$ by $\mathcal{H}^{[\lambda_1\lambda_2\lambda_3]}$. Since a generic field on $b_7$ is also an irrep of $SO(7)$, one can define a function over $b_7$
\be 
Y_I^{[\lambda_1\lambda_2\lambda_3]}\in \mathcal{H}^{[\lambda_1\lambda_2\lambda_3]}
\ee 
which transform in a given $SO(7)$ representation. Every $\Phi^{[\lambda_1\lambda_2\lambda_3]}(z)$ can be expanded as,
\be 
\label{eqn:harmonic}
\Phi^{[\lambda_1\lambda_2\lambda_3]}(z)=\sum_I c_I Y_I^{[\lambda_1\lambda_2\lambda_3]}
\ee 
where, $c_I$'s are the coefficients corresponding to the amplitudes of each harmonic mode. Therefore, one can write the harmonic expansion of a generic field on $b_7$ as,
\be 
\Phi^{[J_1,J_2]}_{[\lambda_1, \lambda_2, \lambda_3]}(x,y)=\sum_I \Phi^{[J_1,J_2]}_I(x) Y^I_{[\lambda_1, \lambda_2, \lambda_3]}(y)
\ee 
\par 
 Coming back to the invariant operators $\boxtimes_y$ on $b_7$ whose most significant property  is that, they act diagonally on the harmonics. Therefore the harmonics are the eigenfunctions of the operator $\boxtimes_y$, i.e
\be 
\boxtimes_y^{[\lambda_1, \lambda_2, \lambda_3]}Y_I^{[\lambda_1, \lambda_2, \lambda_3]}=M(\lambda_1, \lambda_2, \lambda_3)Y_I^{[\lambda_1, \lambda_2, \lambda_3]}\;.
\ee 
This reduces the 11D differential equation \eqref{eqn:sugra_eqn}  to a simpler form as follows,
\be \label{eqn:harmonic_eigenvalue_eqn}
&&\Big(\Box_x^{[J_1,J_2]} +\boxtimes_y^{[\lambda_1, \lambda_2, \lambda_3]}\Big)\sum_I \Phi^{[J_1,J_2]}(x) Y^I_{[\lambda_1, \lambda_2, \lambda_3]}(y)=0\nn\\
&&\implies \Big(\Box_x^{[J_1,J_2]} +M(\lambda_1, \lambda_2, \lambda_3)\Big) \Phi^{[J_1,J_2]}(x)=0
\ee 
which is an algebraic equation for the mass spectrum of the 4 dimensional  fields.
Finally, A generic field that is an irrep of  $SO(3,2)\times U(1)_L\times U(1)_R$  can be expanded as follows,
\be 
\Phi^{[J_1J_2]}_{Z_{L\,charge},Z_{R\,charge}}(x,y)=\sum_{[M_1,M_2,M_3]} \sum_m \mathcal{H}^{[M_1,M_2,M_3]m}_{Z_{L\,charge},Z_{R\,charge}}(y).\,\phi^{[J_1,J_2]}_{[M_1,M_2,M_3]}(x)
\ee 
where the two sums above run over $U(3)$ and $U(1)_L\times U(1)_R$ representations, $Z_{L\,charge},Z_{R\,charge}$ are the charges of $U(1)_L$ and $U(1)_R$ respectively.
 The expansion of any generic field on $\ads\times U(1)_L\setminus  U(3)/U(1)_R$ contains only the harmonics whose $H=U(1)_L\times U(1)_R$ and $U(3)$ quantum numbers are such that the $U(3)$ representation decomposed under $H$, contain the $H$ representation of the field. This fact constraints the $U(3)$ quantum numbers of the associated harmonics.

\subsection{Invariant  operators on $b_7$}
In this section we derive the invariant operators on $b_7$.
The $SO(7)$ covariant derivative  is defined as,
\be 
\mathcal{D}=d+\mathcal{B}^{\alpha\beta}\,(T^{SO(7)})_{\alpha\beta}
\ee 
$(T^{SO(7)})_{\alpha\beta}$ are the generators of $SO(7)$ which take different representations depending on the object $\mathcal{D}$ acts on. For example, 
\begin{itemize}
	\item For $SO(7)$ scalar $(T^{SO(7)})_{\alpha\beta}=0$.
	\item For $SO(7)$ vector $(T^{SO(7)}_{\alpha\beta})_\gamma^{\;\rho}=\eta_{\gamma\epsilon}\,\delta^\epsilon_{[\alpha}\delta^\rho_{\beta]}$.
	\item For $SO(7)$ spinor representation $(T^{SO(7)})_{\alpha\beta}=\frac{1}{4}\tau_{[\alpha}\tau_{\beta]}$.
\end{itemize}
To compute the eigenvalues of the harmonics by solving \eqref{eqn:harmonic_eigenvalue_eqn}, it is necessary to have an explicit form of the $SO(7)$ covariant derivative $\mathcal{D}$. Notably, the operator $\mathcal{D}$ acts differently on various differential forms.
 Since $U(1)$ is a subgroup of $SO(7)$, $\mathcal{D}$ can be written as,\cite{fermionauria}
\be 
\mathcal{D}=d+\Omega_H T_H+ \mathcal{C}\gamma M_\gamma:=\mathcal{D}^H+ \mathcal{C}\gamma M_\gamma
\ee 
Now, consider the following term,
\be 
&&\Omega L^{-1}= L^{-1} dL L^{-1}
=- L^{-1}  L dL^{-1}= -dL^{-1} \implies  dL^{-1}=-\Omega L^{-1}\;.
\ee 
The above is written using the following identity,
\be 
d(L L^{-1})=0\implies dL. L^{-1}+ L dL^{-1} =0\implies dL. L^{-1}=- L dL^{-1}\;.
\ee 
The covariant derivative $\mathcal{D}^H$ can now be expressed as follows,
\be 
\mathcal{D}^H&=&d+\Omega^H T_H\nn\\
\mathcal{D}^H L^{-1}&=&(d+\Omega^H T_H)L^{-1}=dL^{-1}+\Omega^H T_H L^{-1}=-\Omega L^{-1}+\Omega^H T_H L^{-1}\nn\\
&=&-(\Omega^\alpha T_\alpha + \Omega^H T_H) L^{-1}+\Omega^H T_H L^{-1}\nn\\
\mathcal{D}^H&=& \Omega^\alpha T_\alpha\;.
\ee 
Now, in case of spinors the $SO(7)$ covariant derivative is,
\be 
&&d-\frac{1}{4}\mathcal{B}^{\alpha\beta}\tau_{\alpha\beta}\nn\\
&=&d+\frac{1}{4}f^{3AB}\tau_{AB}\,\Omega_{3}+\frac{1}{4}f^{8AB}\tau_{AB}\Omega_{8}+\frac{1}{4}f^{3\dot{A}\dot{B}}\tau_{\dot{A}\dot{B}}\,\Omega_{3}+\frac{1}{4}f^{8\dot{A}\dot{B}}\tau_{\dot{A}\dot{B}}\,\Omega_{8}\nn\\
&-&\frac{1}{4}\Bigg[\Big(\frac{b\dot{b}}{2a}
+\frac{a\dot{b}}{2b} 
-\frac{ab}{2\dot{b}}\Big)f^{mA\dot{B}}\tau_{mA}
+(-\frac{\dot{b}^2}{2c}f^{3\dot{A}\dot{B}}
-\frac{\sqrt{3}\dot{b}^2}{c} f^{8\dot{A}\dot{B}})\tau_{\dot{A}Z}\Bigg]\mathcal{B}^{\dot{B}}\nn\\
&-&\frac{1}{4}\Bigg[\Big(-\frac{b\dot{b}}{2a}
+\frac{a\dot{b}}{2b}
-\frac{ab}{2\dot{b}}\Big)f^{mB\dot{A}}\tau_{m\dot{A}}-(\frac{b^2}{2c}f^{3AB}+\frac{\sqrt{3}b^2}{c}f^{8AB})\tau_{AZ}\Bigg]\mathcal{B}^{B}\nn\\
&-&\frac{1}{4}\Bigg[\frac{a^2}{2c}f^{3nm}\tau_{mZ} 
+\Big(-\frac{b\dot{b}}{2a}
+\frac{a\dot{b}}{2b} 
+\frac{ab}{2\dot{b}}\Big)f^{nA\dot{A}}\tau_{A\dot{A}}\Bigg]\mathcal{B}^n\nn\\
&-&\frac{1}{4} \Bigg[(-\frac{\dot{b}^2}{2c}f^{3\dot{A}\dot{B}}
-\frac{\sqrt{3}\dot{b}^2}{c} f^{8\dot{A}\dot{B}})\tau_{\dot{A}\dot{B}}+(-\frac{b^2}{2c}f_{3AB}-\frac{\sqrt{3}b^2}{c}f^{8AB})\tau_{AB}\Bigg]\mathcal{B}^Z\nn\\
&=& \mathcal{D}^{H}+M_\alpha \mathcal{B}^\alpha\;.
\ee 
where,
\be 
\mathcal{D}^H=d+\frac{1}{4}f^{3AB}\tau_{AB}\,\Omega_{3}-f^{8AB}\tau_{AB}\Omega_{8}\frac{1}{4}f^{3\dot{A}\dot{B}}\tau_{\dot{A}\dot{B}}\,\Omega_{3}+\frac{1}{4}f^{8\dot{A}\dot{B}}\tau_{\dot{A}\dot{B}}\,\Omega_{8}\;,
\ee 
and
\be 
M_\alpha \mathcal{B}^\alpha&=&-\frac{1}{4}\Bigg[\frac{a^2}{2c}f^{3nm}\tau_{mZ} 
+\Big(-\frac{b\dot{b}}{2a}
+\frac{a\dot{b}}{2b} 
+\frac{ab}{2\dot{b}}\Big)f^{nA\dot{A}}\tau_{A\dot{A}}\Bigg]\mathcal{B}^n\nn\\
&-&\frac{1}{4}\Bigg[\Big(\frac{b\dot{b}}{2a}
+\frac{a\dot{b}}{2b} 
-\frac{ab}{2\dot{b}}\Big)f^{mA\dot{B}}\tau_{mA}
+(-\frac{\dot{b}^2}{2c}f^{3\dot{A}\dot{B}}
-\frac{\sqrt{3}\dot{b}^2}{c} f^{8\dot{A}\dot{B}})\tau_{\dot{A}Z}\Bigg]\mathcal{B}^{\dot{B}}\nn\\
&-&\frac{1}{4}\Bigg[\Big(-\frac{b\dot{b}}{2a}
+\frac{a\dot{b}}{2b}
-\frac{ab}{2\dot{b}}\Big)f^{mB\dot{A}}\tau_{m\dot{A}}-(\frac{b^2}{2c}f^{3AB}+\frac{\sqrt{3}b^2}{c}f^{8AB})\tau_{AZ}\Bigg]\mathcal{B}^{B}\nn\\
&-&\frac{1}{4} \Bigg[(-\frac{\dot{b}^2}{2c}f^{3\dot{A}\dot{B}}
-\frac{\sqrt{3}\dot{b}^2}{c} f^{8\dot{A}\dot{B}})\tau_{\dot{A}\dot{B}}+(-\frac{b^2}{2c}f_{3AB}-\frac{\sqrt{3}b^2}{c}f^{8AB})\tau_{AB}\Bigg]\mathcal{B}^Z.
\ee 
From the above equation we obtain the following matrix forms of $M_\alpha$'s,
\be 
&&M_n=\frac{a^2}{2c}f^{3nm}\tau_{mZ} 
+\Big(-\frac{b\dot{b}}{2a}
+\frac{a\dot{b}}{2b} 
+\frac{ab}{2\dot{b}}\Big)f^{nA\dot{A}}\tau_{A\dot{A}}\nn\\
&&\implies M_n^{mZ}=\frac{a^2}{2c}f^{3nm} \qquad
M_n^{{A\dot{A}}}=\Big(-\frac{b\dot{b}}{2a}
+\frac{a\dot{b}}{2b} 
+\frac{ab}{2\dot{b}}\Big)f^{nA\dot{A}}\nn\\
&&\implies M_n^{mZ}=2\sqrt{\frac{2}{3}}f^{3nm} \qquad
M_n^{{A\dot{A}}}=-2\sqrt{6}f^{nA\dot{A}}
\ee

\be 
M_B=\Big(-\frac{b\dot{b}}{2a}
+\frac{a\dot{b}}{2b}
-\frac{ab}{2\dot{b}}\Big)f^{mB\dot{A}}\tau_{m\dot{A}}-(\frac{b^2}{2c}f^{3AB}+\frac{\sqrt{3}b^2}{c}f^{8AB})\tau_{AZ}\nn\\
\implies M_B^{m\dot{A}}=\Big(-\frac{b\dot{b}}{2a}
+\frac{a\dot{b}}{2b}
-\frac{ab}{2\dot{b}}\Big)f^{mB\dot{A}}\qquad M_B^{AZ}=-(\frac{b^2}{2c}f^{3AB}+\frac{\sqrt{3}b^2}{c}f^{8AB})\nn\\
\implies M_B^{m\dot{A}}=2\sqrt{6}f^{mB\dot{A}}\qquad M_B^{AZ}=-(2\sqrt{\frac{2}{3}}f^{3AB}+4\sqrt{2}f^{8AB})
\ee

\be 
&&M_{\dot{B}}=\Big(\frac{b\dot{b}}{2a}
+\frac{a\dot{b}}{2b} 
-\frac{ab}{2\dot{b}}\Big)f^{mA\dot{B}}\tau_{mA}
+(-\frac{\dot{b}^2}{2c}f^{3\dot{A}\dot{B}}
-\frac{\sqrt{3}\dot{b}^2}{c} f^{8\dot{A}\dot{B}})\tau_{\dot{A}Z}\nn\\
&&\implies M_{\dot{B}}^{mA}=\Big(\frac{b\dot{b}}{2a}
+\frac{a\dot{b}}{2b} 
-\frac{ab}{2\dot{b}}\Big)f^{mA\dot{B}}\qquad
M_{\dot{B}}^{\dot{A}Z}=+(-\frac{\dot{b}^2}{2c}f^{3\dot{A}\dot{B}}
-\frac{\sqrt{3}\dot{b}^2}{c} f^{8\dot{A}\dot{B}})\nn\\
&&\implies M_{\dot{B}}^{mA}=-2\sqrt{6}f^{mA\dot{B}}\qquad
M_{\dot{B}}^{\dot{A}Z}=(-2\sqrt{\frac{2}{3}}f^{3\dot{A}\dot{B}}
-4\sqrt{2}f^{8\dot{A}\dot{B}})
\ee

\be 
M_Z=(-\frac{\dot{b}^2}{2c}f^{3\dot{A}\dot{B}}
-\frac{\sqrt{3}\dot{b}^2}{c} f^{8\dot{A}\dot{B}})\tau_{\dot{A}\dot{B}}+(-\frac{b^2}{2c}f_{3AB}-\frac{\sqrt{3}b^2}{c}f^{8AB})\tau_{AB}\nn\\
\implies M_Z^{\dot{A}\dot{B}}=(-\frac{\dot{b}^2}{2c}f^{3\dot{A}\dot{B}}
-\frac{\sqrt{3}\dot{b}^2}{c} f^{8\dot{A}\dot{B}})\qquad M_Z^{AB}=(-\frac{b^2}{2c}f_{3AB}-\frac{\sqrt{3}b^2}{c}f^{8AB})\nn\\
\implies M_Z^{\dot{A}\dot{B}}=(-2\sqrt{\frac{2}{3}}f^{3\dot{A}\dot{B}}
-4\sqrt{2} f^{8\dot{A}\dot{B}})\qquad M_Z^{AB}=-(2\sqrt{\frac{2}{3}}f^{3AB}+4\sqrt{2}f^{8AB})
\ee 

\be \label{eqn:M_matrix}
&&M_1=\begin{pmatrix}
	0&0&0&0&0&0&0\\
	0&0&0&0&0&0&2\sqrt{\frac{2}{3}}\\
	0&0&0&0&0&-\sqrt{6}&0\\
	0&0&0&0&\sqrt{6}&0&0\\
	0&0&0&-\sqrt{6}&0&0&0\\
	0&0&\sqrt{6}&0&0&0&0\\
	0&-2\sqrt{\frac{2}{3}}&0&0&0&0&0\\
\end{pmatrix}\quad M_2=\begin{pmatrix}
	0&0&0&0&0&0&-2\sqrt{\frac{2}{3}}\\
	0&0&0&0&0&0&0\\
	0&0&0&0&-\sqrt{6}&0&0\\				0&0&0&0&0&-\sqrt{6}&0\\
	0&0&\sqrt{6}&0&0&0&0\\
	0&0&0&\sqrt{6}&0&0&0\\
	2\sqrt{\frac{2}{3}}&0&0&0&0&0&0\\
\end{pmatrix}\nn\\
&&M_4=\begin{pmatrix}
	0&0&0&0&0&\sqrt{6}&0\\
	0&0&0&0&\sqrt{6}&0&0\\
	0&0&0&0&0&0&0\\				0&0&0&0&0&0&7\sqrt{\frac{2}{3}}\\
	0&-\sqrt{6}&0&0&0&0&0\\
	-\sqrt{6}&0&0&0&0&0&0\\
	0&0&0&-7\sqrt{\frac{2}{3}}&0&0&0\\
\end{pmatrix}\quad
M_5=\begin{pmatrix}
	0&0&0&0&-\sqrt{6}&0&0\\
	0&0&0&0&0&\sqrt{6}&0\\
	0&0&0&0&0&0&-7\sqrt{\frac{2}{3}}\\	0&0&0&0&0&0&0\\
	\sqrt{6}&0&0&0&0&0&0\\
	0&-\sqrt{6}&0&0&0&0&0\\
	0&0&7\sqrt{\frac{2}{3}}&0&0&0&0\\
\end{pmatrix}\nn\\
&&M_6=\begin{pmatrix}
	0&0&0&\sqrt{6}&0&0&0\\
	0&0&-\sqrt{6}&0&0&0&0\\
	0&\sqrt{6}&0&0&0&0&0\\				-\sqrt{6}&0&0&0&0&0&0\\
	0&0&0&0&0&0&0\\
	0&0&0&0&0&0&5\sqrt{\frac{2}{3}}\\
	0&0&0&0&0&-5\sqrt{\frac{2}{3}}&0\\
\end{pmatrix}\quad M_7=\begin{pmatrix}
	0&0&-\sqrt{6}&0&0&0&0\\
	0&0&0&-\sqrt{6}&0&0&0\\
	\sqrt{6}&0&0&0&0&0&0\\				0&\sqrt{6}&0&0&0&0&0\\
	0&0&0&0&0&0&-5\sqrt{\frac{2}{3}}\\
	0&0&0&0&0&0&0\\
	0&0&0&0&5\sqrt{\frac{2}{3}}&0&0\\
\end{pmatrix}\nn\\
&&M_Z=\begin{pmatrix}
	0&0&0&0&0&0&0\\
	0&0&0&0&0&0&0\\
	0&0&0&-7\sqrt{\frac{2}{3}}&0&0&0\\	0&0&7\sqrt{\frac{2}{3}}&0&0&0&0\\
	0&0&0&0&0&-5\sqrt{\frac{2}{3}}&0\\
	0&0&0&0&5\sqrt{\frac{2}{3}}&0&0\\
	0&0&0&0&0&0&0\\
\end{pmatrix}
\ee 
\underline{\textbf{Laplacian operator:}}
The invariant Laplacian operator $\boxtimes^{[000]}$, with $M_\gamma=0$, can be evaluated as follows,
\be 
\label{eqn:laplacian}
\boxtimes^{[000]}_y:=\eta^{\alpha\beta}\mathcal{D}^{b_7}_\alpha\mathcal{D}^{b_7}_\beta
=-\mathcal{D}^{b_7}_1\mathcal{D}^{b_7}_1-\mathcal{D}^{b_7}_2\mathcal{D}^{b_7}_2-\mathcal{D}^{b_7}_4\mathcal{D}^{b_7}_4-\mathcal{D}^{b_7}_5\mathcal{D}^{b_7}_5-\mathcal{D}^{b_7}_6\mathcal{D}^{b_7}_6-\mathcal{D}^{b_7}_7\mathcal{D}^{b_7}_7-\mathcal{D}^{b_7}_Z\mathcal{D}^{b_7}_Z
\ee 
where, $\mathcal{D}^{b_7}_\alpha$ denotes the projection along the vielbein directions.
Now,
\be 
&&\mathcal{D}^{b_7}=\mathcal{B}^\alpha \mathcal{D}^{b_7}_\alpha=-\Omega^\alpha T_\alpha\nn\\
&&\implies \mathcal{B}^m \mathcal{D}^{b_7}_m+\mathcal{B}^A \mathcal{D}^{b_7}_A+\mathcal{B}^{\dot{A}} \mathcal{D}^{b_7}_{\dot{A}}+\mathcal{B}^Z \mathcal{D}^{b_7}_Z=-\Omega^m T_m-\Omega^A T_A-\Omega^{\dot{A}} T_{\dot{A}}-\Omega^Z T_Z
\ee 
where, $T_\alpha$ are the generators of $b_7$ \eqref{eqn:b7_generator}. Therefore the above equation implies,
\be 
&& \mathcal{D}^{b_7}_m=-4\sqrt{6}\, \lambda_m \qquad \mathcal{D}^{b_7}_A=-4\sqrt{6}\, \lambda_A\qquad \mathcal{D}^{b_7}_{\dot{A}}=4\sqrt{6}\, \lambda_{\dot{A}}\qquad \mathcal{D}^{b_7}_Z=-12\sqrt{26}\, Z.
\ee 
Thus we get,
\be 
\mathcal{D}_\alpha^{b_7}=\Big(-4\sqrt{6}\, \lambda_m-4\sqrt{6}\, \lambda_A+4\sqrt{6}\, \lambda_{\dot{A}}-12\sqrt{26}\, Z\Big)\;.
\ee 
Substituting the above in \eqref{eqn:laplacian} we obtain,
\be 
\boxtimes^{[000]}
=-96\; \lambda_m^2-96\; \lambda_A^2-96\; \lambda_{\dot{A}}^2-3744\; Z^2\;.
\ee 
\underline{\textbf{Hodge de-Rham operator on one forms:}} This operator takes the following form,
\be 
&&\mathcal{D}_\beta=\mathcal{D}_\beta^H+M_\beta\implies 
\mathcal{D}_\alpha\mathcal{D}^\alpha \mathcal{Y}^\rho=\eta^{\alpha\beta}\mathcal{D}_\alpha\mathcal{D}_\beta=\eta^{\alpha\beta}(\mathcal{D}_\alpha^H+M_\alpha)(\mathcal{D}_\beta^H+M_\beta)\nn\\
&&=\mathcal{D}_\alpha^H\mathcal{D}^{\alpha H} \mathcal{Y}^\rho+\eta^{\alpha\beta}\Big(2(M_\alpha)^\gamma_{\;\;\rho}\mathcal{D}_\beta^H+(M_\alpha)^\gamma_{\;\;\epsilon}(M_\beta)^\epsilon_{\;\;\rho} \Big)\mathcal{Y}^\rho \;.
\ee 
Therefore the Hodge de-Rham operator on one form fields in $b_7$ is,
\be 
\label{eqn:oneform_operator}
&&\boxtimes_y^{[1,0,0]}\mathcal{Y}^\alpha
=(\mathcal{D}_\beta^H\mathcal{D}^{\beta H}+24) \mathcal{Y}^\alpha+\eta^{\gamma\delta}\Big(2(M_\gamma)^\alpha_{\;\;\beta}\mathcal{D}_\delta^H+(M_\gamma)^\alpha_{\;\;\epsilon}(M_\delta)^\epsilon_{\;\;\beta} \Big)\mathcal{Y}^\beta\;.
\ee 
We now have all the relevant operators needed to act on the differential forms on $SO(7)$. 

\subsection{Irreducible representations of $U(3)$}
This is an appropriate place to note down the useful properties of the irreducible representations of $U(3)$ that will be used later\cite{hamermesh}. The irreps of $U(3)$ are labeled by three numbers $[\lambda_1\lambda_2 \lambda_3]$, while the irreps of $U(1)_L, U(1)_R$ are labeled by two numbers $Z_{L\,charge},Z_{R\,charge}$ defined in \eqref{eqn:u1charges}. Listed below are some important properties of the 
$U(n)$ irreducible representation.
\begin{itemize}
\item Irreducible representations (irreps) of \( U(n) \) are also irreps of  general linear group \( GL(n) \).
\item In \( U(n) \), there are no additional constraints on the generators, such as tracelessness. Therefore, only symmetric or antisymmetric tensor representations are possible.
	
	\item The above argument leads to the results that irreps of $GL(n)$ remain irrep in $U(n)$. Therefore we list the properties of Young Tableau(YT) of $GL(n)$ below.

		\item Irreducible tensor representations of $GL(n)$ can be represented by a YT.
		\item A Young tableau for $GL(n)$ has a maximum of $n$ rows. The columns are anti-symmetrized while the rows are symmetrized.
		The YT for a rank $r$ tensor is labeled by
	\be 
	[\lambda_1\lambda_2\cdots \lambda_n], \quad \lambda_1+\lambda_2+\cdots \lambda_n=r,\quad \lambda_1\geq \lambda_2\geq\cdots \lambda_n\geq 0
	\ee 
	\be 
	\Yboxdim15pt \young(\cdot\cdot  \cdot\cdots \cdot\cdot\cdot,b\cdot \cdots\cdot\cdot  y,.\cdots\cdot x,.\cdots z)\qquad b,x,...z=1,2,...,r
	\ee 
\item  \underline{\textbf{Generic YT of $U(3)$}:} Going back to our case with $n=3$,
a $U(3)$ Young diagram is as follows,
\be 
&&\Yboxdim15pt \young(1\cdots 1\cdots 1\cdots 2\cdots 1\cdots 2\cdots 3\cdots  ,2\cdots 2\cdots 3\cdots 3\cdots ,3\cdots\empty\empty\empty)\nn\\
&&\underbrace{\hskip 1.2 cm}_{M_3}\underbrace{\hskip 2.9 cm}_{M_2}\underbrace{\hskip 3.5 cm}_{M_1}\nn\\
\ee 
The above is the only possible Young tableau due to the requirement of symmetrized rows and antisymmetrized columns.
The number of boxes labeled by 1,2 or 3 can be determined by partitioning the values $M_1,M_2,M_3$ as follows.
\be 
&&\Yboxdim15pt \young(1\cdots 1\cdots 1\cdots 2\cdots 1\cdots 2\cdots 3\cdots  ,2\cdots 2\cdots 3\cdots 3\cdots ,3\cdots\empty\empty\empty)\nn\\
&&\underbrace{\hskip 1.2 cm}_{M_3}\underbrace{\hskip .9 cm}_{p_1}\underbrace{\hskip .9 cm}_{p_2}\underbrace{\hskip .9 cm}_{p_3}\underbrace{\hskip 1 cm}_{q_1}\underbrace{\hskip 1 cm}_{q_2}\underbrace{\hskip 1 cm}_{q_3}
\ee 
Hence,
\be 
\textit{no. of boxes labelled by 1}=M_3+p_1+p_2+q_1\nn\\
\textit{no. of boxes labelled by 2}=M_3+p_1+p_3+q_2\nn\\
\textit{no. of boxes labelled by 3}=M_3+p_2+p_3+q_3
\ee 
Therefore the harmonics on $B_7$ are specified by $[M_1, M_2, M_3, Z_{L\,charge}, Z_{R\,charge}]$. 
	
	\item  \textbf{Dimensionality of representations:}
	\be 
	\textit{Dimension of rank r symmetric tensor}={n+r-1\choose r}\nn\\
	\textit{Dimension of rank r antisymmetric tensor}={n\choose r}\nn\\
	\textit{Dimension of rank r mixed tensor of type \Yboxdim15pt  \young(ij ,k)}=\frac{n(n^2-1)}{3}
	\ee 
\end{itemize}

Now, let us define the following charges,
\be 
\frac{1}{2}\lambda_3\quad charge=Y_3\nn\\
\sqrt{3}\lambda_8\quad charge=Y_8\nn\\
\frac{\sqrt{3}}{\sqrt{2}}\lambda_9\quad charge=Y_9\nn\\
\ee 
which in turn specify the $Z_{L\,charge}, Z_{R\,charge}$ of a representation as follows,
\be 
\label{eqn:u1charges}
Z_{charge} &=& i Y_3 + i Y_8 + \frac{3i }{2}\;Y_9\nn\\
Z_{L\,charge} &=&  i Y_3 + i Y_8 - \frac{13i  }{9}\; Y_9	\nn\\
Z_{R\,charge} &=& -12 i\; Y_3 + i Y_8 
\ee
The eigenvalues of $\lambda_3,\sqrt{3}\lambda_8, \frac{\sqrt{3}}{\sqrt{2}}\lambda_9$  are
\be
\lambda_3\ket{1}&=&\ket{1},\qquad \lambda_3\ket{2}=-\ket{2},\qquad \lambda_3\ket{3}=0\nn\\
\sqrt{3}\lambda_8\ket{1}&=&\ket{1},\qquad
\sqrt{3}\lambda_8\ket{2}=\ket{2},\qquad
\sqrt{3}\lambda_8\ket{3}=-2\ket{3}\nn\\
\frac{\sqrt{3}}{\sqrt{2}}\lambda_9\ket{1}&=&\ket{1},\qquad\frac{\sqrt{3}}{\sqrt{2}}\lambda_9\ket{2}=\ket{2},\qquad\frac{\sqrt{3}}{\sqrt{2}}\lambda_9\ket{3}=\ket{3}
\ee 
Thus,
\be 
&&Y_3=(M_3+p_1+p_2+q_1)-(M_3+p_1+p_3+q_2)\nn=p_2+q_1-p_3-q_2\\
&&Y_8=(M_3+p_1+p_2+q_1)+(M_3+p_1+p_3+q_2)-2(M_3+p_2+p_3+q_3)\nn\\
&&=2p_1+q_1+q_2-p_2-p_3-2q_3\nn\\
&&Y_9=3M_3+2M_2+M_1\qquad M_2=p_1+p_2+p_3, M_1=q_1+q_2+q_3\nn\\&&=3M_3+2p_1+2p_2+2p_3+q_1+q_2+q_3
\ee 
In terms of the positive integers $M_3, p_1,p_2,p_3,q_1,q_2,q_3$ we have,
\be 
\label{eqn:Z_ev}
Z_{charge} 
&=& i\Big( 2q_1-2p_3 +  2p_1-2q_3 + \frac{9}{2}M_3+3p_1+3p_2+3p_3+\frac{3 }{2}q_1+\frac{3 }{2}q_2+\frac{3 }{2}q_3\Big),
\ee 
\be 
\label{eqn:ZL_charge}
Z_{L\,charge} &=&  i \Big(2q_1-2p_3 +  2p_1-2q_3 - \frac{13 }{9}(3M_3+2p_1+2p_2+2p_3+q_1+q_2+q_3)\Big),\ee 
\be 
\label{eqn:ZR_charge}
Z_{R\,charge}
&=& i\Big(2p_1-13p_2+11p_3-11q_1+13q_2-2q_3 \Big).
\ee

\subsection{Spectra of zero-form fields}
The $SO(7)$ scalar is labelled by $[0,0,0]$ which breaks under $H$ representation as $[0,0,0]$ which implies,
\be 
Z^0_{L\,charge} &=&  i Y_3 + i Y_8 - \frac{13i  }{9}\; Y_9=0\implies	13i Y_3-\frac{13i  }{9}\; Y_9=0\implies  Y_3=\frac{1}{9}\; Y_9 \nn\\
Z^0_{R\,charge} &=& -12 i\; Y_3 + i Y_8 =0\implies Y_8=12Y_3
\ee
Substituting the above in \eqref{eqn:u1charges} we find,
\be 
Z^0_{charge} &=& i( Y_3 +  Y_8 + \frac{3 }{2}Y_9)=i( \frac{1}{9}Y_9 +  \frac{4}{3}Y_9 + \frac{3 }{2}Y_9)=i\frac{53}{18}Y_9
\ee 
Now our task is to evaluate the eigenvalues of $\lambda_m^2,\lambda_A^2,\lambda_{\dot{A}}^2, Z^2$ on a generic $U(3)$ Young tableau.
We find below that the eigenvalue of the operator $\lambda_A^2+\lambda_{\dot{A}}^2$ depends only on $M_1,M_2$. Since the left most $M_3$ boxes remain invariant upon action of the $\lambda_A^2+\lambda_{\dot{A}}^2$ due to antisymmetrized column. Rest of the tableau is the  $SU(3)$ part. Therefore the eigenvalues are those obtained for an $SU(3)$ YT. However, the difference between an $SU(3)$ and $U(3)$ YT is that the latter is labelled by three integers $\{M_1,M_2,M_3\}$. The $U(1)$ charge of a YT is essentially the $\frac{\sqrt{3}}{\sqrt{2}}\lambda_9$ charge. Therefore, the  eigenvalue  of operator  $\lambda_A^2+\lambda_{\dot{A}}^2$ on a generic $U(3)$ YT is
\be 
4\Big(M_1+M_2+M_1M_2\Big)
\ee 
and $\lambda_m^2$ is  
\be 
2\Big(M_1+M_2+2M_1M_2\Big)
\ee 
Thus,
\be 
\boxtimes_y^{[000]}\mathcal{Y}
=\Big(-96\; \lambda_m^2-96\; \lambda_A^2-96\; \lambda_{\dot{A}}^2-3744\; (Z^{0}_{charge})^2\Big)\mathcal{Y}
\ee 
Substituting the appropriate eigenvalues above we find,
\be 
\boxtimes_y^{[000]}\mathcal{Y}
&=&-96\Big(\; \lambda_m^2+ \lambda_A^2+ \lambda_{\dot{A}}^2\Big)\mathcal{Y}-3744\; Z^2\mathcal{Y}\nn\\
&=&-96\Big(2(M_1+M_2+2M_1M_2)+ 4(M_1+M_2+M_1M_2)\Big)\mathcal{Y}-3744\; (Z^{0}_{charge})^2\mathcal{Y}\nn\\
&=&-96\Big(6M_1+6M_2+8M_1M_2\Big)\mathcal{Y}-3744\; (Z^{0}_{charge})^2\mathcal{Y}=H_0 \mathcal{Y}
\ee 
\subsection{Spectra of one-form fields}
Let us consider a one form field of $SO(7)$, which implies
\be 
[\lambda_1, \lambda_2, \lambda_3]=[100]
\ee 
The one form fields are in the vector representation of $SO(7)$ denoted by 
$\mathcal{Y}^\alpha$.
Now, the 7 dimensional vector representation $[1,0,0]$ of $SO(7)$ can be decomposed under $H=U(1)_L\times U(1)_R$ representation as follows.  
\be 
\mathbf{[1,0,0]}\to 7&&=\mathbf{1\oplus\bar{1}\oplus1\oplus\bar{1}\oplus 1\oplus\bar{1}\oplus 1}=\mathbf{1\oplus3\oplus\bar{3}}
\ee 
In the above equation, we have one real irrep and 3 complex irreps under $(U(1)_L, U(1)_R)$.
The RHS represents how the $U(1)$ representations combine to form a $U(3)$ representation. Below, we outline the construction of a three dimensional $U(3)$ representation.
\be 
&&\mathbf{1\oplus3\oplus\bar{3}}\nn\\
&&=[0,0,0]\oplus [100]\oplus [100]=[0,0,0]\oplus [1,1,0]\oplus [1,1,0]
\ee 
where,
$[100]$ is rank $1$ symmetric tensor representation which is 3 dimensional.
$[110]$ is rank $2$ symmetric tensor representation which is also 3 dimensional. The $u(3)$ irreps can be further broken down in terms of $u(2)\times u(1)$ irreps, which is the isometry,labelled by $[J,m,n]$ as follows,
\be 
\mathbf{1\oplus3\oplus\bar{3}}=\mathbf{1\oplus2\oplus\bar{2}\oplus1\oplus\bar{1}}=[0,0,0]\oplus[\frac{1}{2},m,n]\oplus[\frac{1}{2},\bar{m},\bar{n}]\oplus[0,m1,n1]\oplus[0,\bar{m1},\bar{n1}]
\ee 

\par Following the notation of \cite{Fabbri:1999mk}, we decompose the $SO(7)$ one form fields in terms of  $(U(1)_L, U(1)_R)$ below,
\be \label{eqn:one_form_decomposition}
&&\mathcal{Y}^A= \lambda^A_{31}\; \langle 1 |  I\rangle+ \lambda^A_{13}\; \; \langle 1 |  I\rangle^*,\nn\\
&&\mathcal{Y}^{\dot{A}}= \lambda^{\dot{A}}_{32}\; \langle 1 |  I\rangle+ \lambda^{\dot{A}}_{23}\; \; \langle 1 |  I\rangle^*,\nn\\
&&\mathcal{Y}^m= \lambda^m_{21} \; \langle 1 |  I\rangle+ \lambda^m_{12} \; \langle 1 |  I\rangle^*,\nn\\
&&\mathcal{Y}^Z= [1 |  I],
\ee 
where,
\be \label{eqn:distinct_eigenvectors}
\langle 1 |  I\rangle&=&\mathcal{H}^{[{Z_L}^1_{ charge},{Z_R}^1_{ charge}]}.W_{\ads}\\
\langle 1 |  I\rangle^*&=&\mathcal{H}^{[\bar{Z}^1_{L\;charge},\bar{Z}^1_{R\;charge}]}.\widetilde{W}_{\ads}\\
\left[1|I\right]&=&\mathcal{H}^{[0,0]}.W_{\ads}.
\ee 
All the quantities in the angular or square bracket above are $SO(3,2)\times U(1)_L\times U(1)_R$ irrreducible fields on $\ads\times B_7$.
$\langle\; | \; \rangle$ denotes a complex  field transforming under $U(1)$ while $\left[1|I\right]$ is real. ${Z_L}^1_{ charge},{Z_R}^1_{ charge}$ are given in \eqref{eqn:ZL_charge} and \eqref{eqn:ZR_charge}. Expanding \eqref{eqn:one_form_decomposition} we find
\be
&&\mathcal{Y}^4=  \langle 1 |  I\rangle+  \langle 1 |  I\rangle^*, \qquad
\mathcal{Y}^5= i\Big(\; \langle 1 |  I\rangle-\; \; \langle 1 |  I\rangle^*\Big),\qquad
\mathcal{Y}^{6}=  \langle 1 |  I\rangle+ \langle 1 |  I\rangle^*,\nn\\
&&\mathcal{Y}^{7}= i\Big(\; \langle 1 |  I\rangle-\; \; \langle 1 |  I\rangle^*\Big),\qquad
\mathcal{Y}^1= \langle 1 |  I\rangle+  \; \langle 1 |  I\rangle^*,\qquad
\mathcal{Y}^2=i\Big(  \langle 1 |  I\rangle - \; \langle 1 |  I\rangle^*\Big),\nn\\
&&\mathcal{Y}^Z= [1 |  I].
\ee 
At this stage, one applies the Hodge de Rham operator in \eqref{eqn:oneform_operator} on $\mathcal{Y}^\alpha$ to find its mass spectrum. Consider the case when $\alpha=4$. We have the following eigenvalue equation
\be 
&&\boxtimes_y^{[1,0,0]}\mathcal{Y}^4 
=(\mathcal{D}_\beta^H\mathcal{D}^{\beta H}+24) \mathcal{Y}^4+\eta^{\gamma\delta}\Big(2(M_\gamma)^4_{\;\;\beta}\mathcal{D}_\delta^H+(M_\gamma)^4_{\;\;\epsilon}(M_\delta)^\epsilon_{\;\;\beta} \Big)\mathcal{Y}^\beta
\ee 
Note that, to visualize the equation, we recall that $\mathcal{H}^{[{Z_L}^1_{ charge},{Z_R}^1_{ charge}]}$ is an YT labelled by some integer numbers, i.e specifying a representation. Hence the operator $\mathcal{D}_\beta^H\mathcal{D}^{\beta H}$ acts on $\mathcal{H}^{[{Z_L}^1_{ charge},{Z_R}^1_{ charge}]}$, and produces eigenvalues depending on those integer numbers. The matrix operator $\boxtimes_y^{[1,0,0]}$ can be computed from above using \eqref{eqn:M_matrix}, which takes the following form,
\be 
\left(
\begin{array}{ccccccc}
	\mathcal{D}^2+\frac{152}{3} & 0 & 2 \sqrt{6} \mathcal{D}_6 & -2 \sqrt{6} \mathcal{D}_5 & 2 \sqrt{6} \mathcal{D}_4 & -2
	\sqrt{6} \mathcal{D}_3 & 4 \sqrt{\frac{2}{3}} \mathcal{D}_2 \\
	0 & \mathcal{D}^2+\frac{152}{3} & 2 \sqrt{6} \mathcal{D}_5 & 2 \sqrt{6} \mathcal{D}_6 & -2 \sqrt{6} \mathcal{D}_3 & -2
	\sqrt{6} \mathcal{D}_4 & -4 \sqrt{\frac{2}{3}} \mathcal{D}_1 \\
	-2 \sqrt{6} \mathcal{D}_6 & -2 \sqrt{6} \mathcal{D}_5 & \mathcal{D}^2+\frac{340}{3} & 14 \sqrt{\frac{2}{3}} \mathcal{D}_7 &
	2 \sqrt{6} \mathcal{D}_2 & 2 \sqrt{6} \mathcal{D}_1 & 14 \sqrt{\frac{2}{3}} \mathcal{D}_4 \\
	2 \sqrt{6} \mathcal{D}_5 & -2 \sqrt{6} \mathcal{D}_6 & -14 \sqrt{\frac{2}{3}} \mathcal{D}_7 & \mathcal{D}^2+\frac{340}{3} &
	-2 \sqrt{6} \mathcal{D}_1 & 2 \sqrt{6} \mathcal{D}_2 & -14 \sqrt{\frac{2}{3}} \mathcal{D}_3 \\
	-2 \sqrt{6} \mathcal{D}_4 & 2 \sqrt{6} \mathcal{D}_3 & -2 \sqrt{6} \mathcal{D}_2 & 2 \sqrt{6} \mathcal{D}_1 &
	\mathcal{D}^2+\frac{244}{3} & 10 \sqrt{\frac{2}{3}} \mathcal{D}_7 & 10 \sqrt{\frac{2}{3}} \mathcal{D}_6 \\
	2 \sqrt{6} \mathcal{D}_3 & 2 \sqrt{6} \mathcal{D}_4 & -2 \sqrt{6} \mathcal{D}_1 & -2 \sqrt{6} \mathcal{D}_2 & -10
	\sqrt{\frac{2}{3}} \mathcal{D}_7 & \mathcal{D}^2+\frac{244}{3} & -10 \sqrt{\frac{2}{3}} \mathcal{D}_5 \\
	-4 \sqrt{\frac{2}{3}} \mathcal{D}_2 & 4 \sqrt{\frac{2}{3}} \mathcal{D}_1 & -14 \sqrt{\frac{2}{3}} \mathcal{D}_4 & 14
	\sqrt{\frac{2}{3}} \mathcal{D}_3 & -10 \sqrt{\frac{2}{3}} \mathcal{D}_6 & 10 \sqrt{\frac{2}{3}} \mathcal{D}_5 &
	\mathcal{D}^2+128 \\
\end{array}
\right)
\ee 
However, as we have three distinct eigenvectors from \eqref{eqn:distinct_eigenvectors} the above matrix can be cast into a $3\times 3$ matrix for transverse fields $\mathcal{D}_\alpha\mathcal{Y}^\alpha=0$ as follows,
\be 
\begin{pmatrix}
	\frac{224K_1}{H_1}&0& 2\sqrt{84K_1}\\	0&18H_1-\frac{224K_1}{H_1}+4\sqrt{14K_1}& 0\\	2\sqrt{14K_1}&0& \sqrt{\frac{3}{8}}H_1
\end{pmatrix}\qquad K_1=\Big(M_1+M_2+M_1M_2\Big)
\ee 
where $H_1$ is the eigenvalue of the operator $\mathcal{D}^2$,
\be 
&&(\mathcal{D}_\beta^H\mathcal{D}^{\beta H}+24) \mathcal{Y}^4\nn\\
&&=\Big(-96\; \lambda_m^2-96\; \lambda_A^2-96\; \lambda_{\dot{A}}^2-3744\; Z^2+24\Big)\mathcal{H}^{[{Z_L}^1_{ charge},{Z_R}^1_{ charge}]}.W_{\ads}\nn\\
&&+\Big(-96\; \lambda_m^2-96\; \lambda_A^2-96\; \lambda_{\dot{A}}^2-3744\; Z^2+24\Big) \mathcal{H}^{[\bar{Z}^1_{L\;charge},\bar{Z}^1_{R\;charge}]} .\widetilde{W}_{\ads}\nn\\ [3mm]
&&=-\Big[96\big(6M_1+6M_2+8M_1M_2\big)+3744\; Z_{charge}^2\Big]\mathcal{H}^{[{Z_L}^1_{ charge},{Z_R}^1_{ charge}]}.W_{\ads}+24 \mathcal{H}^{[{Z_L}^1_{ charge},{Z_R}^1_{ charge}]}.W_{\ads}\nn\\
&&-\Big[96\big(6M_1+6M_2+8M_1M_2\big)+3744\; Z_{charge}^2\Big] \mathcal{H}^{[\bar{Z}^1_{L\;charge},\bar{Z}^1_{R\;charge}]} .W_{\ads}+24 \mathcal{H}^{[\bar{Z}^1_{L\;charge},\bar{Z}^1_{R\;charge}]} .\widetilde{W}_{\ads}\nn\\ [3mm]
&&=\Big(H_1+24\Big)\mathcal{H}^{[{Z_L}^1_{ charge},{Z_R}^1_{ charge}]}.W_{\ads}
+\Big(H_1+24\Big) \mathcal{H}^{[\bar{Z}^1_{L\;charge},\bar{Z}^1_{R\;charge}]} .\widetilde{W}_{\ads}\nn\\
&&=(H_1+24)\mathcal{Y}^4.
\ee 
%The remaining terms in the eigenvalue equation gives,\be 
%&&\eta^{\gamma\delta}\Big(2(M_\gamma)^4_{\;\;\beta}\mathcal{D}_\delta^H+(M_\gamma)^4_{\;\;\epsilon}(M_\delta)^\epsilon_{\;\;\beta} \Big)\mathcal{Y}^\beta\nn\\[2mm]&&=2\sqrt{6}\mathcal{D}_1^H\mathcal{Y}^7+6 \mathcal{Y}^4+2\sqrt{6}\mathcal{D}_2^H\mathcal{Y}^6+6 \mathcal{Y}^4+14\sqrt{\frac{2}{3}}\mathcal{D}_5^H\mathcal{Y}^Z-\frac{98}{3}\mathcal{Y}^4+2\sqrt{6}\mathcal{D}_6^H\mathcal{Y}^2-6 \mathcal{Y}^4-2\sqrt{6}\mathcal{D}_7^H\mathcal{Y}^1\nn\\&&+6\mathcal{Y}^4+14\sqrt{\frac{2}{3}}\mathcal{D}_Z^H\mathcal{Y}^5+\frac{98}{3}\mathcal{Y}^4 \nn\\ [2mm]&&=2\sqrt{6}\mathcal{D}_1^H\mathcal{Y}^7+2\sqrt{6}\mathcal{D}_2^H\mathcal{Y}^6+14\sqrt{\frac{2}{3}}\mathcal{D}_5^H\mathcal{Y}^Z+2\sqrt{6}\mathcal{D}_6^H\mathcal{Y}^2-2\sqrt{6}\mathcal{D}_7^H\mathcal{Y}^1+14\sqrt{\frac{2}{3}}\mathcal{D}_Z^H\mathcal{Y}^5+12\mathcal{Y}^4\nn\\\ee 
%For transverse fields we have $\mathcal{D}_\alpha\mathcal{Y}^\alpha=0$. Therefore, for transverse fields
%This completes solving the eigenvalue equation of the transverse 1-form fields.
One can explicitly find $H_1$ by substituting appropriate values of $[{Z_L}^1_{ charge},{Z_R}^1_{ charge}]$.
\section{Conclusion and outlook}\label{sec:conclusion}

In this paper we studied the compactification of 11-dimensional supergravity on the space \( \text{AdS}_4 \times B_7 \). We examined how fields defined in 11-dimensional spacetime can be expanded in terms of harmonic functions on \( B_7 \). These harmonics are labelled by  quantum numbers specified by the internal manifold $B_7$. Most importantly,
these harmonics are eigenstates of invariant operators of $B_7$, such as the Laplace-Beltrami operator, Hodge de Rham operator etc. 
\par  In this paper we have used the group $SO(7)$ extensively. The reason is,
$SO(7)$ being the rotation group in 7 dimensions preserves the flat metric. Therefore the local symmetries of the compact internal manifold $B_7$ can be understood as being embedded in $SO(7)$. Even though the isometry group of $B_7$ is smaller $SO(7)$ provides a useful framework for harmonic analysis and representation theory. 

\par In this work, we have determined the mass spectrum of \( SO(7) \) scalar and transverse vector fields using the techniques of harmonic analysis on the non-homogeneous Einstein manifold $B_7$.
\par  However, to obtain the complete mass spectrum for the $\ads\times B_7$ compactification, it is necessary to solve the eigenvalue equations for spinor and two-form fields. The eigenvalues of higher-form fields can then be expressed in terms of these, together with the zero- and one-form fields computed in this study. Following this, the spectra must be organized into $OSp(3|4)$ multiplets. This would provide the full information required to compute the superconformal index in the corresponding dual gravity theory, which is the primary objective of this program.

	\begin{center}
		\textbf{Acknowledgments}
	\end{center}
I am deeply grateful to Ashoke Sen for his invaluable guidance, for explaining crucial concepts, and for many insightful discussions. I would also like to extend my heartfelt thanks to Sudhakar Panda, Kamal Panigrahi, Jyotirmoy Bhattacharya, Sayantani Bhattacharyya, Shiraz Minwalla, Palash Dubey, Nibedita Padhi, Sourav Singha and Aditya Sharma for their helpful discussions.
This work was supported by the National Post Doctoral Fellowship, file no. PDF/2022/000876, funded by the Government of India.
	\appendix
	\section{Notations and conventions}\label{app:A}
	\label{appendix:A}
	The representation of $u(3)\cong su(3)\times u(1)$ that we have used are as follows,
	\be 
	\lambda_1=\begin{pmatrix}
		0&1&0\\
		1&0&0\\
		0&0&0
	\end{pmatrix},\quad	\lambda_2=\begin{pmatrix}
		0&-i&0\\
		i&0&0\\
		0&0&0
	\end{pmatrix},\quad 
	\lambda_3=\begin{pmatrix}
		1&0&0\\
		0&-1&0\\
		0&0&0
	\end{pmatrix},\nn\\
	\lambda_4=\begin{pmatrix}
		0&0&1\\
		0&0&0\\
		1&0&0
	\end{pmatrix},\quad	\lambda_5=\begin{pmatrix}
		0&0&-i\\
		0&0&0\\
		i&0&0
	\end{pmatrix}, \quad
	\lambda_6=\begin{pmatrix}
		0&0&0\\
		0&0&1\\
		0&1&0
	\end{pmatrix},\nn\\
	\lambda_7=\begin{pmatrix}
		0&0&0\\
		0&0&-i\\
		0&i&0
	\end{pmatrix},\quad 	\lambda_8=\frac{1}{\sqrt{3}}\begin{pmatrix}
		1&0&0\\
		0&1&0\\
		0&0&-2
	\end{pmatrix},\qquad \lambda_9=\frac{\sqrt{2}}{\sqrt{3}}\begin{pmatrix}
	1&0&0\\
	0&1&0\\
	0&0&1
	\end{pmatrix}.
	\ee 
where, the $su(3)$ generators are the standard Gell-Mann matrices. We also use the following normalization,
	\be 
	\Tr(\lambda_i\lambda_j)=2\delta_{ij}
	\ee 
The structure constants are given by,
\be 
f^{123}=1,\quad f^{147}=f^{156}=f^{246}=f^{257}=f^{345}=f^{367}=\frac{1}{2},\quad f^{458}=f^{678}=\frac{\sqrt{3}}{2}, f^{9xy}=0, x,y=1,...8\nn\\
\ee
The $SO(7)$ Clifford algebra elements are,
\be 
&&\tau_1=\begin{pmatrix}
	0&0&0&0&-i&0&0&0\\
	0&0&0&0&0&-i&0&0\\
	0&0&0&0&0&0&-i&0\\
	0&0&0&0&0&0&0&-i\\
	-i&0&0&0&0&0&0&0\\
	0&-i&0&0&0&0&0&0\\
	0&0&-i&0&0&0&0&0\\
	0&0&0&-i&0&0&0&0
\end{pmatrix}\quad \tau_2=\begin{pmatrix}
0&0&0&0&-1&0&0&0\\
0&0&0&0&0&-1&0&0\\
0&0&0&0&0&0&-1&0\\
0&0&0&0&0&0&0&-1\\
1&0&0&0&0&0&0&0\\
0&1&0&0&0&0&0&0\\
0&0&1&0&0&0&0&0\\
0&0&0&1&0&0&0&0
\end{pmatrix}\nn\\
&&\tau_3=\begin{pmatrix}
	i&0&0&0&0&0&0&0\\
	0&i&0&0&0&0&0&0\\
	0&0&-i&0&0&0&0&0\\
	0&0&0&-i&0&0&0&0\\
	0&0&0&0&-i&0&0&0\\
	0&0&0&0&0&-i&0&0\\
	0&0&0&0&0&0&i&0\\
	0&0&0&0&0&0&0&i
\end{pmatrix}\quad \tau_4=\begin{pmatrix}
0&0&0&i&0&0&0&0\\
0&0&i&0&0&0&0&0\\
0&i&0&0&0&0&0&0\\
i&0&0&0&0&0&0&0\\
0&0&0&0&0&0&0&-i\\
0&0&0&0&0&0&-i&0\\
0&0&0&0&0&-i&0&0\\
0&0&0&0&-i&0&0&0
\end{pmatrix}
\ee 
\be 
&&\tau_5=\begin{pmatrix}
	0&0&0&1&0&0&0&0\\
	0&0&-1&0&0&0&0&0\\
	0&1&0&0&0&0&0&0\\
	-1&0&0&0&0&0&0&0\\
	0&0&0&0&0&0&0&-1\\
	0&0&0&0&0&0&1&0\\
	0&0&0&0&0&-1&0&0\\
	0&0&0&0&1&0&0&0
\end{pmatrix}\quad \tau_6=\begin{pmatrix}
0&0&-i&0&0&0&0&0\\
0&0&0&i&0&0&0&0\\
-i&0&0&0&0&0&0&0\\
0&i&0&0&0&0&0&0\\
0&0&0&0&0&0&i&0\\
0&0&0&0&0&0&0&-i\\
0&0&0&0&i&0&0&0\\
0&0&0&0&0&-i&0&0
\end{pmatrix}\nn\\
&&\tau_7=\begin{pmatrix}
	0&0&1&0&0&0&0&0\\
	0&0&0&1&0&0&0&0\\
	-1&0&0&0&0&0&0&0\\
	0&-1&0&0&0&0&0&0\\
	0&0&0&0&0&0&-1&0\\
	0&0&0&0&0&0&0&-1\\
	0&0&0&0&1&0&0&0\\
	0&0&0&0&0&1&0&0
\end{pmatrix}.
\ee 

\section{Ricci tensor computation}\label{appendix:B}
To evaluate the spin connection $\mathcal{B}^{\alpha\beta}$
we first impose torsion free condition on $b_7$ through the following equation,
\be \label{eqn:no_torsion1}
d \mathcal{B}^\alpha-\mathcal{B}^{\alpha\beta}\wedge \mathcal{B}_\beta=0
\ee
Since we do not know the explicit metric on the manifold, we perform the computation in non-coordinate basis. First we realize the $\mathcal{B}^{\alpha\beta}$'s are one form and can be written in terms of the vielbeins as follows,
\be 
\mathcal{B}^{AB}&=&\mathcal{B}^{AB}_{(C)}\,\mathcal{B}^C+\mathcal{B}^{AB}_{(\dot{C})}\,\mathcal{B}^{\dot{C}}+\mathcal{B}^{AB}_{(m)}\,\mathcal{B}^m+\mathcal{B}^{AB}_{(Z)}\,\mathcal{B}^Z\qquad \mathcal{B}^{A\dot{B}}=\mathcal{B}^{A\dot{B}}_{(C)}\,\mathcal{B}^C+\mathcal{B}^{A\dot{B}}_{(\dot{C})}\,\mathcal{B}^{\dot{C}}+\mathcal{B}^{A\dot{B}}_{(m)}\,\mathcal{B}^m+\mathcal{B}^{A\dot{B}}_{(Z)}\,\mathcal{B}^Z\nn\\
\mathcal{B}^{Am}&=&\mathcal{B}^{Am}_{(C)}\,\mathcal{B}^C+\mathcal{B}^{Am}_{(\dot{C})}\,\mathcal{B}^{\dot{C}}+\mathcal{B}^{Am}_{(n)}\,\mathcal{B}^n+\mathcal{B}^{Am}_{(Z)}\,\mathcal{B}^Z\qquad \mathcal{B}^{\dot{A}m}=\mathcal{B}^{\dot{A}m}_{(C)}\,\mathcal{B}^C+\mathcal{B}^{\dot{A}m}_{(\dot{C})}\,\mathcal{B}^{\dot{C}}+\mathcal{B}^{\dot{A}m}_{(n)}\,\mathcal{B}^n+\mathcal{B}^{\dot{A}m}_{(Z)}\,\mathcal{B}^Z\nn\\
\mathcal{B}^{AZ}&=&\mathcal{B}^{AZ}_{(C)}\,\mathcal{B}^C+\mathcal{B}^{AZ}_{(\dot{C})}\,\mathcal{B}^{\dot{C}}+\mathcal{B}^{AZ}_{(m)}\,\mathcal{B}^m+\mathcal{B}^{AZ}_{(Z)}\,\mathcal{B}^Z\quad \mathcal{B}^{\dot{A}Z}=\mathcal{B}^{\dot{A}Z}_{(C)}\,\mathcal{B}^C+\mathcal{B}^{\dot{A}Z}_{(\dot{C})}\,\mathcal{B}^{\dot{C}}+\mathcal{B}^{\dot{A}Z}_{(m)}\,\mathcal{B}^m+\mathcal{B}^{\dot{A}Z}_{(Z)}\,\mathcal{B}^Z\nn\\
\mathcal{B}^{mn}&=&\mathcal{B}^{mn}_{(C)}\,\mathcal{B}^C+\mathcal{B}^{mn}_{(\dot{C})}\,\mathcal{B}^{\dot{C}}+\mathcal{B}^{mn}_{(p)}\,\mathcal{B}^p+\mathcal{B}^{mn}_{(Z)}\,\mathcal{B}^Z\qquad
\mathcal{B}^{mZ}=\mathcal{B}^{mZ}_{(C)}\,\mathcal{B}^C+\mathcal{B}^{mZ}_{(\dot{C})}\,\mathcal{B}^{\dot{C}}+\mathcal{B}^{mZ}_{(n)}\,\mathcal{B}^n+\mathcal{B}^{mZ}_{(Z)}\,\mathcal{B}^Z\nn\\
\mathcal{B}^{\dot{A}\dot{B}}&=&\mathcal{B}^{\dot{A}\dot{B}}_{(C)}\,\mathcal{B}^C+\mathcal{B}^{\dot{A}\dot{B}}_{(\dot{C})}\,\mathcal{B}^{\dot{C}}+\mathcal{B}^{\dot{A}\dot{B}}_{(m)}\,\mathcal{B}^m+\mathcal{B}^{\dot{A}\dot{B}}_{(Z)}\,\mathcal{B}^Z.
\ee 
Now, note that the indices are raised and lowered by $SO(7)$ metric $\eta^{\alpha\beta}=(-,-,-,-,-,-,-)$ as follows,
\be 
\mathcal{B}^{AB}_{(C)}\,\Omega^C=\mathcal{B}^{AB}_{(C)}\,\eta^{CD}\Omega_D=-\mathcal{B}^{AB}_{(C)}\,\delta^{CD}\Omega_D
=-\mathcal{B}^{AB}_{(D)}\,\Omega_D.
\ee 
Next, we expand the no torsion condition \eqref{eqn:no_torsion1} for $\alpha=m,A,\dot{A},Z$ below. \begin{itemize}
	\item $\alpha=m$
	\be 
	&&	d \mathcal{B}^m-\mathcal{B}^{m\beta}\wedge \mathcal{B}_\beta=0 \implies d \mathcal{B}^m-\mathcal{B}^{mA}\wedge \mathcal{B}_A-\mathcal{B}^{m\dot{A}}\wedge \mathcal{B}_{\dot{A}}-\mathcal{B}^{mn}\wedge \mathcal{B}_n-\mathcal{B}^{mZ}\wedge \mathcal{B}_Z=0\nn\\
	&&	 \implies \frac{1}{a} d {\Omega}^m-\Big(\mathcal{B}^{mA}_{(C)}\,\mathcal{B}^C+\mathcal{B}^{mA}_{(\dot{A})}\mathcal{B}^{\dot{A}}+\mathcal{B}^{mA}_{(m)}\,\mathcal{B}^m+\mathcal{B}^{mA}_{(Z)}\,\mathcal{B}^Z\Big)\wedge \mathcal{B}_A\nn\\
	&&-\Big(\mathcal{B}^{m\dot{A}}_{(C)}\,\mathcal{B}^C+\mathcal{B}^{m\dot{A}}_{(\dot{C})}\,\mathcal{B}^{\dot{C}}+\mathcal{B}^{m\dot{A}}_{(n)}\,\mathcal{B}^n+\mathcal{B}^{m\dot{A}}_{(Z)}\,\mathcal{B}^Z\Big)\wedge \mathcal{B}_{\dot{A}}\nn\\
	&&-\Big(\mathcal{B}^{mn}_{(C)}\,\mathcal{B}^C++\mathcal{B}^{mn}_{(\dot{A})}\mathcal{B}^{\dot{A}}+\mathcal{B}^{mn}_{(p)}\,\mathcal{B}^p+\mathcal{B}^{mn}_{(Z)}\,\mathcal{B}^Z\Big)\wedge \mathcal{B}_n\nn\\
	&&-\Big(\mathcal{B}^{mZ}_{(C)}\,\mathcal{B}^C+\mathcal{B}^{mZ}_{(m)}\,\mathcal{B}^m+\mathcal{B}^{mZ}_{(\dot{A})}\mathcal{B}^{\dot{A}}\Big)\wedge \mathcal{B}_Z=0\nn\\[ 3mm]
	&&	 	 \implies  (f_{mn3}\,\mathcal{B}^n\wedge \Omega^3+\frac{b\dot{b}}{a}f_{mA\dot{B}}\, \mathcal{B}^A\wedge \mathcal{B}^{\dot{B}})\nn\\
	&&-\mathcal{B}^{mA}_{(C)}\mathcal{B}^C\wedge \mathcal{B}_A-\mathcal{B}^{mA}_{(\dot{A})}\mathcal{B}^{\dot{A}}\wedge \mathcal{B}_A-\mathcal{B}^{mA}_{(m)}\mathcal{B}^m\wedge \mathcal{B}_A-\mathcal{B}^{mA}_{(Z)}\mathcal{B}^Z\wedge \mathcal{B}_A\nn\\
	&&-\mathcal{B}^{m\dot{A}}_{(C)}\,\mathcal{B}^C\wedge \mathcal{B}_{\dot{A}}-\mathcal{B}^{m\dot{A}}_{(\dot{C})}\,\mathcal{B}^{\dot{C}}\wedge \mathcal{B}_{\dot{A}}-\mathcal{B}^{m\dot{A}}_{(n)}\,\mathcal{B}^n\wedge \mathcal{B}_{\dot{A}}-\mathcal{B}^{m\dot{A}}_{(Z)}\mathcal{B}^Z\wedge \mathcal{B}_{\dot{A}}\nn\\
	&&-\mathcal{B}^{mn}_{(C)}\,\mathcal{B}^C\wedge \mathcal{B}_n-\mathcal{B}^{mn}_{(\dot{A})}\mathcal{B}^{\dot{A}}\wedge \mathcal{B}_n-\mathcal{B}^{mn}_{(p)}\,\mathcal{B}^p\wedge \mathcal{B}_n-\mathcal{B}^{mn}_{(Z)}\,\mathcal{B}^Z\wedge \mathcal{B}_n\nn\\
	&&-\mathcal{B}^{mZ}_{(C)}\,\mathcal{B}^C\wedge\mathcal{B}_Z-\mathcal{B}^{mZ}_{(m)}\,\mathcal{B}^m\wedge \mathcal{B}_Z-\mathcal{B}^{mZ}_{(\dot{A})}\mathcal{B}^{\dot{A}}\wedge \mathcal{B}_Z=0
	\ee 
Equating coeffcients of each independent basis components	of above equation, we find the following relations,
	\be  \label{eqn:alpha_m}
\hspace*{-1.5cm}	-\mathcal{B}^{mA}_{(n)}\mathcal{B}_A\wedge\mathcal{B}_n
	-\mathcal{B}^{m\dot{A}}_{(n)} \mathcal{B}_{\dot{A}}\wedge \mathcal{B}_n-\mathcal{B}^{mZ}_{(n)} \mathcal{B}_Z\wedge \mathcal{B}_n
-(\mathcal{B}^{mn}_{(C)}\,\mathcal{B}^C+\mathcal{B}^{mn}_{(\dot{A})}\mathcal{B}^{\dot{A}}+\mathcal{B}^{mn}_{(p)}\mathcal{B}^p+\mathcal{B}^{mn}_{(Z)}\mathcal{B}^Z)\wedge \mathcal{B}_n&=&f_{mn3} \Omega^3\wedge \mathcal{B}^n\nn\\
 (\frac{b\dot{b}}{a}f_{mA\dot{B}}\, \mathcal{B}^A\wedge \mathcal{B}^{\dot{B}})
	-\mathcal{B}^{mA}_{(\dot{A})}\mathcal{B}^{\dot{A}}\wedge \mathcal{B}_A
	-\mathcal{B}^{m\dot{A}}_{(C)}\,\mathcal{B}^C\wedge \mathcal{B}_{\dot{A}}&=&0\nn\\	 	 
	\mathcal{B}^{mA}_{(C)}=0\quad \mathcal{B}^{m\dot{A}}_{(\dot{C})}=0
\qquad \mathcal{B}^{m\dot{A}}_{(Z)} =\mathcal{B}^{mZ}_{(\dot{A})}\qquad
\mathcal{B}^{mZ}_{(A)}&=&\mathcal{B}^{mA}_{(Z)}
	\ee 
	
	\item   $\alpha=A$
	\be 
	&&	d \mathcal{B}^A-\mathcal{B}^{A\beta}\wedge \mathcal{B}_\beta=0\implies d \mathcal{B}^A+\mathcal{B}^{mA}\wedge \mathcal{B}_m-\mathcal{B}^{AB}\wedge \mathcal{B}_B-\mathcal{B}^{AZ}\wedge \mathcal{B}_Z-\mathcal{B}^{A\dot{B}}\wedge \mathcal{B}_{\dot{B}}=0\nn\\ [3mm]
	&&	 \implies -\frac{a\dot{b}}{b} f^{mA\dot{B}}\mathcal{B}_{m}\wedge \mathcal{B}_{\dot{B}}-f^{3AB}\,\Omega_{3}\wedge \mathcal{B}_B-f^{8AB}\Omega_{8}\wedge \mathcal{B}_B\nn\\
	&&+\mathcal{B}^{mA}_{(\dot{A})}\mathcal{B}^{\dot{A}}\wedge \mathcal{B}_m-\mathcal{B}^{A\dot{B}}_{(m)}\,\mathcal{B}^m\wedge \mathcal{B}_{\dot{B}}+\mathcal{B}^{mA}_{(n)}\,\mathcal{B}^n\wedge \mathcal{B}_m+\mathcal{B}^{mA}_{(Z)}\,\mathcal{B}^Z\wedge \mathcal{B}_m\nn\\
	&&-\mathcal{B}^{AB}_{(C)}\,\mathcal{B}^C\wedge \mathcal{B}_B-\mathcal{B}^{AB}_{(\dot{C})}\,\mathcal{B}^{\dot{C}}\wedge \mathcal{B}_B-\mathcal{B}^{AB}_{(p)}\,\mathcal{B}^p\wedge \mathcal{B}_B-\mathcal{B}^{AB}_{(Z)}\,\mathcal{B}^Z\wedge \mathcal{B}_B\nn\\
	&&-\mathcal{B}^{AZ}_{(C)}\,\mathcal{B}^C\wedge \mathcal{B}_Z-\mathcal{B}^{AZ}_{(\dot{A})}\,\mathcal{B}^{\dot{C}}\wedge \mathcal{B}_Z-\mathcal{B}^{AZ}_{(m)}\,\mathcal{B}^m\wedge \mathcal{B}_Z\nn\\
	&&-\mathcal{B}^{A\dot{B}}_{(C)}\,\mathcal{B}^C\wedge \mathcal{B}_{\dot{B}}-\mathcal{B}^{A\dot{B}}_{(\dot{C})}\,\mathcal{B}^{\dot{C}}\wedge \mathcal{B}_{\dot{B}}-\mathcal{B}^{A\dot{B}}_{(Z)}\,\mathcal{B}^Z\wedge \mathcal{B}_{\dot{B}}+\mathcal{B}^{mA}_{(C)}\mathcal{B}^C\wedge \mathcal{B}_m=0
	\ee
	The above equation leads to,
	\begin{align}\label{eqn:alpha_A}
				&	 -\frac{a\dot{b}}{b} f^{mA\dot{B}}\mathcal{B}_{m}\wedge \mathcal{B}_{\dot{B}}
		+\mathcal{B}^{mA}_{(\dot{A})}\mathcal{B}^{\dot{A}}\wedge \mathcal{B}_m-\mathcal{B}^{A\dot{B}}_{(m)}\,\mathcal{B}^m\wedge \mathcal{B}_{\dot{B}}=0&\nn\\[2mm]
		&	-f^{3AB}\,\Omega_{3}\wedge\mathcal{B}_B-f^{8AB}\Omega_{8}\wedge \mathcal{B}_B
	-\mathcal{B}^{AB}_{(C)}\,\mathcal{B}^C\wedge \mathcal{B}_B-\mathcal{B}^{AB}_{(\dot{C})}\,\mathcal{B}^{\dot{C}}\wedge \mathcal{B}_B-\mathcal{B}^{AB}_{(p)}\,\mathcal{B}^p\wedge \mathcal{B}_B-\mathcal{B}^{AB}_{(Z)}\,\mathcal{B}^Z\wedge \mathcal{B}_B
	&\nn\\&-\mathcal{B}^{AZ}_{(C)}\,\mathcal{B}^C\wedge \mathcal{B}_Z
	-\mathcal{B}^{A\dot{B}}_{(C)}\,\mathcal{B}^C\wedge \mathcal{B}_{\dot{B}}=0&\nn\\[2mm]
		& 
	\mathcal{B}^{mA}_{(n)}=0,\quad \mathcal{B}^{mA}_{(Z)} =-\mathcal{B}^{AZ}_{(m)},\quad \mathcal{B}^{AZ}_{(\dot{B})} =\mathcal{B}^{A\dot{B}}_{(Z)},\quad \mathcal{B}^{A\dot{B}}_{(\dot{C})}=0&
\end{align}	 
	\item  $\alpha=\dot{A}$
	\be 
	&&	d \mathcal{B}^{\dot{A}}-\mathcal{B}^{\dot{A}\beta}\wedge \mathcal{B}_\beta=0\implies d \mathcal{B}^{\dot{A}}+\mathcal{B}^{m\dot{A}}\wedge \mathcal{B}_m-\mathcal{B}^{\dot{A}B}\wedge \mathcal{B}_B-\mathcal{B}^{\dot{A}Z}\wedge \mathcal{B}_Z-\mathcal{B}^{\dot{A}\dot{B}}\wedge \mathcal{B}_{\dot{B}}=0\nn\\
	&&	 \implies \frac{1}{\dot{b}} d {\Omega}^{\dot{A}}+\Big(\mathcal{B}^{m\dot{A}}_{(C)}\mathcal{B}^C+\mathcal{B}^{m\dot{A}}_{(n)}\,\mathcal{B}^n+\mathcal{B}^{m\dot{A}}_{(Z)}\,\mathcal{B}^Z\Big)\wedge \mathcal{B}_m\nn\\
	&&-\Big(\mathcal{B}^{\dot{A}B}_{(C)}\,\mathcal{B}^C+\mathcal{B}^{\dot{A}B}_{(p)}\,\mathcal{B}^p+\mathcal{B}^{\dot{A}B}_{(Z)}\,\mathcal{B}^Z\Big)\wedge \mathcal{B}_B-\Big(\mathcal{B}^{\dot{A}Z}_{(C)}\,\mathcal{B}^C+\mathcal{B}^{\dot{A}Z}_{(\dot{A})}\,\mathcal{B}^{\dot{C}}+\mathcal{B}^{\dot{A}Z}_{(m)}\,\mathcal{B}^m\Big)\wedge \mathcal{B}_Z\nn\\
	&&-\Big(\mathcal{B}^{\dot{A}\dot{B}}_{(C)}\,\mathcal{B}^C+\mathcal{B}^{\dot{A}\dot{B}}_{(\dot{C})}\,\mathcal{B}^{\dot{C}}+\mathcal{B}^{\dot{A}\dot{B}}_{(m)}\,\mathcal{B}^m+\mathcal{B}^{\dot{A}\dot{B}}_{(Z)}\,\mathcal{B}^Z\Big)\wedge \mathcal{B}_{\dot{B}}=0\nn\\ [3mm]
	&&	 \implies  -\frac{ab}{\dot{b}}f^{m\dot{A}{B}}\mathcal{B}_{m}\wedge \mathcal{B}_{{B}}-f^{3\dot{A}\dot{B}}\,\Omega_{3}\wedge \mathcal{B}_{\dot{B}}-f^{8\dot{A}\dot{B}}\,\Omega_{8}\wedge \mathcal{B}_{\dot{B}}\nn\\
	&&+\mathcal{B}^{m\dot{A}}_{(C)}\mathcal{B}^C\wedge \mathcal{B}_m+\mathcal{B}^{m\dot{A}}_{(n)}\,\mathcal{B}^n\wedge \mathcal{B}_m+\mathcal{B}^{m\dot{A}}_{(Z)}\mathcal{B}^Z\wedge \mathcal{B}_m-\mathcal{B}^{\dot{A}B}_{(C)}\,\mathcal{B}^C\wedge \mathcal{B}_B-\mathcal{B}^{\dot{A}B}_{(p)}\,\mathcal{B}^p\wedge \mathcal{B}_B-\mathcal{B}^{\dot{A}B}_{(Z)}\,\mathcal{B}^Z\wedge \mathcal{B}_B\nn\\
	&&-\Big(\mathcal{B}^{\dot{A}Z}_{(C)}\,\mathcal{B}^C+\mathcal{B}^{\dot{A}Z}_{(\dot{A})}\,\mathcal{B}^{\dot{C}}+\mathcal{B}^{\dot{A}Z}_{(m)}\,\mathcal{B}^m\Big)\wedge \mathcal{B}_Z-\Big(\mathcal{B}^{\dot{A}\dot{B}}_{(C)}\,\mathcal{B}^C+\mathcal{B}^{\dot{A}\dot{B}}_{(\dot{C})}\,\mathcal{B}^{\dot{C}}+\mathcal{B}^{\dot{A}\dot{B}}_{(m)}\,\mathcal{B}^m+\mathcal{B}^{\dot{A}\dot{B}}_{(Z)}\,\mathcal{B}^Z\Big)\wedge \mathcal{B}_{\dot{B}}=0\nn\\
	\ee
From the above equation we find,
	\begin{align} \label{eqn:alpha_dotA}
	  &-\frac{ab}{\dot{b}}f^{m\dot{A}{B}}\mathcal{B}_{m}\wedge \mathcal{B}_{{B}}+\mathcal{B}^{m\dot{A}}_{(C)}\mathcal{B}^C\wedge \mathcal{B}_m-\mathcal{B}^{\dot{A}B}_{(p)}\,\mathcal{B}^p\wedge \mathcal{B}_B=0&\nn\\
&	\hspace*{-2cm}	 -f^{3\dot{A}\dot{B}}\,\Omega_{3}\wedge \mathcal{B}_{\dot{B}}-f^{8\dot{A}\dot{B}}\,\Omega_{8}\wedge \mathcal{B}_{\dot{B}}
	-\mathcal{B}^{\dot{A}Z}_{(\dot{A})}\,\mathcal{B}^{\dot{C}}\wedge \mathcal{B}_Z
	-\Big(\mathcal{B}^{\dot{A}\dot{B}}_{(C)}\,\mathcal{B}^C+\mathcal{B}^{\dot{A}\dot{B}}_{(\dot{C})}\,\mathcal{B}^{\dot{C}}+\mathcal{B}^{\dot{A}\dot{B}}_{(m)}\,\mathcal{B}^m+\mathcal{B}^{\dot{A}\dot{B}}_{(Z)}\,\mathcal{B}^Z\Big)\wedge \mathcal{B}_{\dot{B}}=0& \nn\\	
&	\mathcal{B}^{m\dot{A}}_{(n)}=0 \qquad \mathcal{B}^{m\dot{A}}_{(Z)}=-\mathcal{B}^{\dot{A}Z}_{(m)}\qquad
	\mathcal{B}^{\dot{A}B}_{(C)}=0 \qquad
	\mathcal{B}^{\dot{A}Z}_{(B)} =\mathcal{B}^{\dot{A}B}_{(Z)}&.
	\end{align}

	\item  $\alpha=Z$
	\be 
	&&	d \mathcal{B}^Z-\mathcal{B}^{Z\beta}\wedge \mathcal{B}_\beta=0\implies d \mathcal{B}^Z-\mathcal{B}^{Zm}\wedge \mathcal{B}_m-\mathcal{B}^{ZA}\wedge \mathcal{B}_A-\mathcal{B}^{Z\dot{A}}\wedge \mathcal{B}_{\dot{A}}=0\nn\\
	&&\hspace*{-2cm}\implies \frac{1}{c} d {\Omega}^Z+	\Big(\mathcal{B}^{mZ}_{(C)}\,\mathcal{B}^C+\mathcal{B}^{mZ}_{(\dot{C})}\,\mathcal{B}^{\dot{C}}+\mathcal{B}^{mZ}_{(p)}\,\mathcal{B}^p+\mathcal{B}^{mZ}_{(Z)}\,\mathcal{B}^Z\Big)\wedge \mathcal{B}_m\nn\\
	&&\hspace*{-2cm}+\Big(\mathcal{B}^{AZ}_{(C)}\,\mathcal{B}^C+\mathcal{B}^{{A}Z}_{(\dot{C})}\,\mathcal{B}^{\dot{C}}+\mathcal{B}^{AZ}_{(m)}\,\mathcal{B}^m+\mathcal{B}^{AZ}_{(Z)}\,\mathcal{B}^Z\Big)\wedge \mathcal{B}_A+\Big(\mathcal{B}^{\dot{A}Z}_{(C)}\,\mathcal{B}^C+\mathcal{B}^{\dot{A}Z}_{(\dot{C})}\,\mathcal{B}^{\dot{C}}+\mathcal{B}^{\dot{A}Z}_{(m)}\,\mathcal{B}^m+\mathcal{B}^{\dot{A}Z}_{(Z)}\,\mathcal{B}^Z\Big)\wedge \mathcal{B}_{\dot{A}}
	=0\nn\\ [2mm]
	&&\hspace*{-2cm}\implies \frac{a^2}{2c}f_{3mn} \mathcal{B}^m\wedge \mathcal{B}^n+(\frac{b^2}{2c}f_{3AB}+\frac{\sqrt{3}b^2}{c}f^{8AB})\mathcal{B}_{A}\wedge \mathcal{B}_B+(\frac{\dot{b}^2}{2c}f_{3\dot{A}\dot{B}}
	+\frac{\sqrt{3}\dot{b}^2}{c} f^{8\dot{A}\dot{B}})\mathcal{B}_{\dot{A}}\wedge \mathcal{B}_{\dot{B}}\nn\\
	&&\hspace*{-2cm}+\Big(\mathcal{B}^{mC}_{(Z)}\,\mathcal{B}^C+\mathcal{B}^{m\dot{C}}_{(Z)}\,\mathcal{B}^{\dot{C}}+\mathcal{B}^{mZ}_{(p)}\,\mathcal{B}^p+\mathcal{B}^{mZ}_{(Z)}\,\mathcal{B}^Z\Big)\wedge \mathcal{B}_m
+\Big(\mathcal{B}^{AZ}_{(C)}\,\mathcal{B}^C+\mathcal{B}^{{A}\dot{C}}_{(Z)}\,\mathcal{B}^{\dot{C}}-\mathcal{B}^{mA}_{(Z)}\,\mathcal{B}^m+\mathcal{B}^{AZ}_{(Z)}\,\mathcal{B}^Z\Big)\wedge \mathcal{B}_A\nn\\
	&&\hspace*{-2cm}+\Big(\mathcal{B}^{\dot{A}C}_{(Z)}\,\mathcal{B}^C+\mathcal{B}^{\dot{A}Z}_{(\dot{C})}\,\mathcal{B}^{\dot{C}}-\mathcal{B}^{m\dot{A}}_{(Z)}\,\mathcal{B}^m+\mathcal{B}^{\dot{A}Z}_{(Z)}\,\mathcal{B}^Z\Big)\wedge \mathcal{B}_{\dot{A}}
	=0
	\ee 
The above results in the following relations
	\be 
	&&\implies \frac{a^2}{2c}f_{3mn} \mathcal{B}^m\wedge \mathcal{B}^n
	+\Big(\mathcal{B}^{mZ}_{(p)}\,\mathcal{B}^p\Big)\wedge \mathcal{B}_m
	=0
	\ee 
	%%%%%%%%%%%%%%%%%%%%
	\be 
	&&\implies (\frac{b^2}{2c}f_{3AB}+\frac{\sqrt{3}b^2}{c}f^{8AB})\mathcal{B}_{A}\wedge \mathcal{B}_B+\Big(\mathcal{B}^{AZ}_{(C)}\,\mathcal{B}^C\Big)\wedge \mathcal{B}_A
	=0
	\ee 
	%%%%%%%%%%%%%%%%%%%%%%%%%%%%%%%%%%
	\be 
	&&\implies (\frac{\dot{b}^2}{2c}f_{3\dot{A}\dot{B}}
	+\frac{\sqrt{3}\dot{b}^2}{c} f^{8\dot{A}\dot{B}})\mathcal{B}_{\dot{A}}\wedge \mathcal{B}_{\dot{B}}+\Big(\mathcal{B}^{\dot{A}Z}_{(\dot{C})}\,\mathcal{B}^{\dot{C}}\Big)\wedge \mathcal{B}_{\dot{A}}
	=0
	\ee 
	%%%%%%%%%%%%%%%%%%
	\be 
	&&\implies 
	\mathcal{B}^{m\dot{A}}_{(Z)}=0,\quad\mathcal{B}^{{A}\dot{C}}_{(Z)}=0,\quad\mathcal{B}^{\dot{A}C}_{(Z)}=0,\mathcal{B}^{mZ}_{(Z)}=0\quad \mathcal{B}^{\dot{A}Z}_{(Z)}=0\quad \mathcal{B}^{AZ}_{(Z)}=0
	\ee 
	\be 
	&&-\mathcal{B}^{mA}_{(Z)}=\mathcal{B}^{mA}_{(Z)}\implies \mathcal{B}^{mA}_{(Z)}=0
	\ee

\end{itemize}	
Combining all the constraints we got above we find many components of the one form spin connection vanishes, which we list below:
\begin{itemize}
	\item  
	\be 
	&&	 	
	\mathcal{B}^{mA}_{(C)}=0\qquad
	\mathcal{B}^{mA}_{(Z)}=0\qquad
	\mathcal{B}^{mA}_{(n)}=0
	\ee 
	\item 
	\be 
	&&	 	
	\mathcal{B}^{m\dot{A}}_{(\dot{C})}=0\qquad
	\mathcal{B}^{m\dot{A}}_{(Z)}=0\qquad
	\mathcal{B}^{m\dot{A}}_{(n)}=0 
	\ee 
	
	\item 
	\be 
	&&	 	
	\mathcal{B}^{mZ}_{(\dot{A})}=0\qquad
	\mathcal{B}^{mZ}_{(A)}=0\qquad\mathcal{B}^{mZ}_{(Z)}=0
	\ee 
	\item 	\be 
	&&
	\mathcal{B}^{{A}\dot{C}}_{(Z)}=0\qquad\mathcal{B}^{\dot{A}B}_{(C)}=0 \qquad
	\mathcal{B}^{A\dot{B}}_{(\dot{C})}=0
	\ee 
	\item 
	\be 
	&& \mathcal{B}^{AZ}_{(Z)}=0\qquad \mathcal{B}^{AZ}_{(m)}=0\qquad\mathcal{B}^{AZ}_{(\dot{B})} =0
	\ee 
	\item 
	\be 
	&& \mathcal{B}^{\dot{A}Z}_{(Z)}=0\qquad	 \mathcal{B}^{\dot{A}Z}_{(m)}=0\qquad
	\mathcal{B}^{\dot{A}Z}_{(B)} =0
	\ee 
\end{itemize}
Substituting the above relations, the non-vanishing components of the $\mathcal{B}^{\alpha\beta}$'s are computed below.
\begin{itemize}	
	\item $\alpha=m$:
From \eqref{eqn:alpha_m} we obtain
	\be 
	&&	 	 \implies  
	-\mathcal{B}^{mZ}_{(n)} \mathcal{B}_Z\wedge \mathcal{B}_n-\mathcal{B}^{mn}\wedge \mathcal{B}_n=f_{mn3} \Omega^3\wedge \mathcal{B}^n\nn\\
	&&	 	 \implies  
	\mathcal{B}^{mn}\wedge \mathcal{B}_n=f_{mn3} \Omega^3\wedge \mathcal{B}_n-\mathcal{B}^{mZ}_{(n)} \mathcal{B}_Z\wedge \mathcal{B}_n\nn\\
	&&	 	 \implies  
	\mathcal{B}^{mn}\wedge \mathcal{B}_n=f_{mn3} \Omega^3\wedge \mathcal{B}_n-\frac{a^2}{2c}f^{3nm} \mathcal{B}_Z\wedge \mathcal{B}_n\nn\\
	&&	 	 \implies  
	\mathcal{B}^{mn}=f_{mn3} \Omega^3-\frac{a^2}{2c}f^{3nm} \mathcal{B}_Z
	\ee 
	
	%%%%%%%%%%%%%%%%%%%%%%%%%%%%%%%%%%%%%%%%%%%%
	\be 
	&&	 	 \implies  (\frac{b\dot{b}}{a}f_{mA\dot{B}}\, \mathcal{B}^A\wedge \mathcal{B}^{\dot{B}})
	-\mathcal{B}^{mA}_{(\dot{A})}\mathcal{B}^{\dot{A}}\wedge \mathcal{B}_A
	-\mathcal{B}^{m\dot{A}}_{(C)}\,\mathcal{B}^C\wedge \mathcal{B}_{\dot{A}}=0
	\ee 		
	
	%%%%%%%%%%%%%%%%%%%%%%%%%%%%%%%%%%%%%%%%%
	
	\item  $\alpha=A$: From \eqref{eqn:alpha_A} we find,
	\be \label{a}
	&&	 \implies -\frac{a\dot{b}}{b} f^{mA\dot{B}}\mathcal{B}_{m}\wedge \mathcal{B}_{\dot{B}}
	+\mathcal{B}^{mA}_{(\dot{A})}\mathcal{B}^{\dot{A}}\wedge \mathcal{B}_m-\mathcal{B}^{A\dot{B}}_{(m)}\,\mathcal{B}^m\wedge \mathcal{B}_{\dot{B}}=0,
	\ee
		\be 
	&&-\mathcal{B}^{AB}\wedge \mathcal{B}_B-\mathcal{B}^{AZ}_{(C)}\,\mathcal{B}^C\wedge \mathcal{B}_Z=f^{3AB}\,\Omega_{3}\wedge \mathcal{B}_B+f^{8AB}\Omega_{8}\wedge \mathcal{B}_B\nn\\
	&&	 \implies -\mathcal{B}^{AB}\wedge \mathcal{B}_B=\Big[-(\frac{b^2}{2c}f_{3AB}+\frac{\sqrt{3}b^2}{c}f^{8AB})\mathcal{B}^{B}\Big]\wedge \mathcal{B}_Z+f^{3AB}\,\Omega_{3}\wedge \mathcal{B}_B+f^{8AB}\Omega_{8}\wedge \mathcal{B}_B\nn\\
	&&	 \implies \mathcal{B}^{AB}=-(\frac{b^2}{2c}f_{3AB}+\frac{\sqrt{3}b^2}{c}f^{8AB})\mathcal{B}^Z -f^{3AB}\,\Omega_{3}-f^{8AB}\Omega_{8}\;.
	\ee
	
	\item  $\alpha=\dot{A}$: \eqref{eqn:alpha_dotA}  gives
	\be \label{adot}
	&&	 \implies  -\frac{ab}{\dot{b}}f^{m\dot{A}{B}}\mathcal{B}_{m}\wedge \mathcal{B}_{{B}}+\mathcal{B}^{m\dot{A}}_{(C)}\mathcal{B}^C\wedge \mathcal{B}_m-\mathcal{B}^{\dot{A}B}_{(p)}\,\mathcal{B}^p\wedge \mathcal{B}_B=0,\nn 
	\ee
	\be 
	&&	 \implies 
	-\mathcal{B}^{\dot{A}\dot{B}}\wedge \mathcal{B}_{\dot{B}}= f^{3\dot{A}\dot{B}}\,\Omega_{3}\wedge \mathcal{B}_{\dot{B}}+f^{8\dot{A}\dot{B}}\,\Omega_{8}\wedge \mathcal{B}_{\dot{B}}
	+\mathcal{B}^{\dot{A}Z}_{(\dot{C})}\,\mathcal{B}^{\dot{C}}\wedge \mathcal{B}_Z\nn\\
	&&	 \implies 
	\mathcal{B}^{\dot{A}\dot{B}}=- f^{3\dot{A}\dot{B}}\,\Omega_{3}-f^{8\dot{A}\dot{B}}\,\Omega_{8}
	+(\frac{\dot{b}^2}{2c}f_{3\dot{A}\dot{B}}
	+\frac{\sqrt{3}\dot{b}^2}{c} f^{8\dot{A}\dot{B}}) \mathcal{B}_Z\;.
	\ee
	
	\item  $\alpha=Z$: In this case we find,
	\be 
	&&\implies \frac{a^2}{2c}f_{3mn} \mathcal{B}^m\wedge \mathcal{B}^n
	+\mathcal{B}^{mZ}_{(n)}\,\mathcal{B}^n\wedge \mathcal{B}_m=0\nn\\
	&&\implies 
	\mathcal{B}^{mZ}_{(n)}\,\mathcal{B}^n=\frac{a^2}{2c}f^{3nm} \mathcal{B}^n
	\ee 
	%%%%%%%%%%%%%%%%%%%%
	\be 
	&&\implies (\frac{b^2}{2c}f_{3AB}+\frac{\sqrt{3}b^2}{c}f^{8AB})\mathcal{B}_{A}\wedge \mathcal{B}_B+\Big(\mathcal{B}^{AZ}_{(C)}\,\mathcal{B}^C\Big)\wedge \mathcal{B}_A=0\nn\\
	&&\implies \Big(\mathcal{B}^{AZ}_{(B)}\,\mathcal{B}^B\Big)\wedge \mathcal{B}_A=-(\frac{b^2}{2c}f_{3AB}+\frac{\sqrt{3}b^2}{c}f^{8AB})\mathcal{B}^{B}\wedge \mathcal{B}_A\nn\\
	&&\implies \mathcal{B}^{AZ}_{(B)}\,\mathcal{B}^B=-(\frac{b^2}{2c}f_{3AB}+\frac{\sqrt{3}b^2}{c}f^{8AB})\mathcal{B}^{B}
	\ee 
	%%%%%%%%%%%%%%%%%%%%%%%%%%%%%%%%%%
	\be 
	&&\implies (\frac{\dot{b}^2}{2c}f_{3\dot{A}\dot{B}}
	+\frac{\sqrt{3}\dot{b}^2}{c} f^{8\dot{A}\dot{B}})\mathcal{B}_{\dot{A}}\wedge \mathcal{B}_{\dot{B}}+\Big(\mathcal{B}^{\dot{A}Z}_{(\dot{C})}\,\mathcal{B}^{\dot{C}}\Big)\wedge \mathcal{B}_{\dot{A}}=0\nn\\
	&&\implies \Big(\mathcal{B}^{\dot{A}Z}_{(\dot{C})}\,\mathcal{B}^{\dot{C}}\Big)\wedge \mathcal{B}_{\dot{A}}=(\frac{\dot{b}^2}{2c}f_{3\dot{A}\dot{B}}
	+\frac{\sqrt{3}\dot{b}^2}{c} f^{8\dot{A}\dot{B}})\mathcal{B}_{\dot{B}}\wedge \mathcal{B}_{\dot{A}}\nn\\
	&&\implies \Big(\mathcal{B}^{\dot{A}Z}_{(\dot{B})}\,\mathcal{B}^{\dot{B}}\Big)=-(\frac{\dot{b}^2}{2c}f_{3\dot{A}\dot{B}}
	+\frac{\sqrt{3}\dot{b}^2}{c} f^{8\dot{A}\dot{B}})\mathcal{B}^{\dot{B}}
	\ee 
	\item 
	\be \label{m}
	&&	 	 (\frac{b\dot{b}}{a}f_{mA\dot{B}}\, \mathcal{B}^A\wedge \mathcal{B}^{\dot{B}})
	-\mathcal{B}^{mA}_{(\dot{A})}\mathcal{B}^{\dot{A}}\wedge \mathcal{B}_A
	-\mathcal{B}^{m\dot{A}}_{(C)}\,\mathcal{B}^C\wedge \mathcal{B}_{\dot{A}}=0
	\ee
	From \eqref{a},\eqref{adot} we find
	\be 
	&&	 -\frac{a\dot{b}}{b} f^{mA\dot{B}}\mathcal{B}_{m}\wedge \mathcal{B}_{\dot{B}}
	+\mathcal{B}^{mA}_{(\dot{A})}\mathcal{B}^{\dot{A}}\wedge \mathcal{B}_m-\mathcal{B}^{A\dot{A}}_{(m)}\,\mathcal{B}^m\wedge \mathcal{B}_{\dot{A}}=0\nn\\
	&&\implies \mathcal{B}^{mA}_{(\dot{A})}\mathcal{B}^{\dot{A}}=\frac{a\dot{b}}{b} f^{mA\dot{B}}\mathcal{B}^{\dot{B}}
	-\mathcal{B}^{A\dot{A}}_{(m)}\mathcal{B}^{\dot{A}}
	\ee
	\be 
	&&	 \mathcal{B}^{m\dot{A}}_{(C)}\mathcal{B}^C\wedge \mathcal{B}_m-\frac{ab}{\dot{b}}f^{m\dot{A}{B}}\mathcal{B}_{m}\wedge \mathcal{B}_{{B}}+\mathcal{B}^{A\dot{A}}_{(m)}\,\mathcal{B}^m\wedge \mathcal{B}_A=0\nn \\
	&&\implies 	 \mathcal{B}^{m\dot{A}}_{(C)}\mathcal{B}^C\wedge \mathcal{B}_m=-\frac{ab}{\dot{b}}f^{m\dot{A}{B}} \mathcal{B}_{{B}}\wedge \mathcal{B}_{m}-\mathcal{B}^{A\dot{A}}_{(m)}\mathcal{B}_A\wedge\mathcal{B}_m \nn \\
	&&\implies 	 \mathcal{B}^{m\dot{A}}_{(A)}\mathcal{B}^A=\frac{ab}{\dot{b}}f^{m\dot{A}{B}} \mathcal{B}^{{B}}+\mathcal{B}^{A\dot{A}}_{(m)}\mathcal{B}^A\;. \nn
	\ee
	Substituting the above in \eqref{m} we obtain
	\be
	&&	 	 (\frac{b\dot{b}}{a}f_{mA\dot{B}}\, \mathcal{B}^A\wedge \mathcal{B}^{\dot{B}})
	-\mathcal{B}^{mA}_{(\dot{A})}\mathcal{B}^{\dot{A}}\wedge \mathcal{B}_A
	-\mathcal{B}^{m\dot{A}}_{(C)}\,\mathcal{B}^C\wedge \mathcal{B}_{\dot{A}}=0\nn\\
	&&\implies 	 	 (\frac{b\dot{b}}{a}f_{mA\dot{B}}\, \mathcal{B}^A\wedge \mathcal{B}^{\dot{B}})
	-\Big[\frac{a\dot{b}}{b} f^{mA\dot{B}}\mathcal{B}^{\dot{B}}
	-\mathcal{B}^{A\dot{A}}_{(m)}\mathcal{B}^{\dot{A}}\Big]\wedge \mathcal{B}_A
	-\Big[\frac{ab}{\dot{b}}f^{m\dot{A}{B}} \mathcal{B}^{{B}}+\mathcal{B}^{A\dot{A}}_{(m)}\mathcal{B}^A\Big]\wedge \mathcal{B}_{\dot{A}}=0\nn\\
	&&\implies 	 	 (\frac{b\dot{b}}{a}f_{mA\dot{B}}\, \mathcal{B}^A\wedge \mathcal{B}^{\dot{B}})
	-\frac{a\dot{b}}{b} f^{mA\dot{B}}\mathcal{B}^A\wedge\mathcal{B}^{\dot{B}} 
	+\frac{ab}{\dot{b}}f^{m\dot{B}{A}} \mathcal{B}^{{A}}\wedge \mathcal{B}^{\dot{B}}+\mathcal{B}^{A\dot{B}}_{(m)}\mathcal{B}^A\wedge \mathcal{B}^{\dot{B}}+\mathcal{B}^{A\dot{B}}_{(m)}\mathcal{B}^A\wedge\mathcal{B}^{\dot{B}}=0\nn\\
	&&\implies 	 	 {\boxed{\mathcal{B}^{A\dot{B}}_{(m)}=-\frac{b\dot{b}}{2a}f_{mA\dot{B}}
			+\frac{a\dot{b}}{2b} f^{mA\dot{B}}
			-\frac{ab}{2\dot{b}}f^{m\dot{B}{A}}}}\;.
	\ee
	
	Thus we find,
	\be 
	&& \mathcal{B}^{mA}_{(\dot{A})}\mathcal{B}^{\dot{A}}=\frac{a\dot{b}}{b} f^{mA\dot{B}}\mathcal{B}^{\dot{B}}
	-\mathcal{B}^{A\dot{B}}_{(m)}\mathcal{B}^{\dot{B}}\nn\\
	&&\implies \mathcal{B}^{mA}_{(\dot{A})}\mathcal{B}^{\dot{A}}=\frac{a\dot{b}}{b} f^{mA\dot{B}}\mathcal{B}^{\dot{B}}
	-\Big[-\frac{b\dot{b}}{2a}f_{mA\dot{B}}
	+\frac{a\dot{b}}{2b} f^{mA\dot{B}}
	-\frac{ab}{2\dot{b}}f^{m\dot{B}{A}}\Big]\mathcal{B}^{\dot{B}}\nn\\
	&&\implies \mathcal{B}^{mA}_{(\dot{B})}\mathcal{B}^{\dot{B}}=
	+\Big[\frac{b\dot{b}}{2a}f_{mA\dot{B}}
	+\frac{a\dot{b}}{2b} f^{mA\dot{B}}
	+\frac{ab}{2\dot{b}}f^{m\dot{B}{A}}\Big]\mathcal{B}^{\dot{B}}\;,
	\ee
and
	\be 
	&& \mathcal{B}^{m\dot{B}}_{(A)}\mathcal{B}^A=\frac{ab}{\dot{b}}f^{m\dot{B}{A}} \mathcal{B}^{{A}}+\mathcal{B}^{A\dot{B}}_{(m)}\mathcal{B}^A \nn\\
	&&\implies \mathcal{B}^{m\dot{B}}_{(A)}\mathcal{B}^A=\frac{ab}{\dot{b}}f^{m\dot{B}{A}} \mathcal{B}^{{A}}+\Big[-\frac{b\dot{b}}{2a}f_{mA\dot{B}}
	+\frac{a\dot{b}}{2b} f^{mA\dot{B}}
	-\frac{ab}{2\dot{b}}f^{m\dot{B}{A}}\Big]\mathcal{B}^A \nn\\
	&&\implies \mathcal{B}^{m\dot{B}}_{(A)}\mathcal{B}^A=\Big[-\frac{b\dot{b}}{2a}f_{mA\dot{B}}
	-\frac{a\dot{b}}{2b} f^{m\dot{B}A}
	+\frac{ab}{2\dot{b}}f^{m\dot{B}{A}}\Big]\mathcal{B}^A\;.
	\ee
\end{itemize}
At this stage we have computed all the components of the spin connection $\mathcal{B}^{\alpha\beta}$ whose final expressions are given in \eqref{eqn:spin_connection}.

The next step is to compute the	 Riemann curvature tensor,
	\be \label{eqref:rieman_curvature}
\mathcal{R}^{\alpha\beta}=d\mathcal{B}^{\alpha\beta}-\mathcal{B}^{\alpha\gamma}\wedge \mathcal{B}^\beta_\gamma:=\half  \mathcal{R}^{\alpha\beta}_{\gamma\delta}\, \mathcal{B}^\gamma\wedge \mathcal{B}^\delta\;.
\ee 
We present the computation of one component of the Riemann tensor and the rest of them can be computed similarly.	For $\alpha,\beta=A,B$ in \eqref{eqref:rieman_curvature} we get,
\be \label{eqn:riemann_components}
\mathcal{R}^{AB}&=&\mathcal{R}^{AB}_{\gamma\delta}\,\mathcal{B}^\gamma\wedge \mathcal{B}^\delta\nn\\
&=&\mathcal{R}^{AB}_{CD}\,\mathcal{B}^C\wedge \mathcal{B}^D+\mathcal{R}^{AB}_{Cm}\,\mathcal{B}^C\wedge \mathcal{B}^m+\mathcal{R}^{AB}_{CZ}\,\mathcal{B}^C\wedge \mathcal{B}^Z+\mathcal{R}^{AB}_{mn}\,\mathcal{B}^m\wedge \mathcal{B}^n+\mathcal{R}^{AB}_{mZ}\,\mathcal{B}^m\wedge \mathcal{B}^Z\;.\nn
\ee 
We also have,
\be 
&&\mathcal{R}^{AB}=d\mathcal{B}^{AB}+\mathcal{B}^{A\sigma}\wedge \mathcal{B}^{\sigma B}\nn\\
&&=d\mathcal{B}^{AB}+\mathcal{B}^{A\dot{C}}\wedge \mathcal{B}^{\dot{C} B}+\mathcal{B}^{AZ}\wedge \mathcal{B}^{Z B}-\mathcal{B}^{mA}\wedge \mathcal{B}^{m B}\nn\\ [3mm]
&&= -\frac{a^2}{2}f^{3AB} f_{3mn}\mathcal{B}^m\wedge \mathcal{B}^n-(\frac{b^2}{2}f^{3AB}f_{3CD}+\frac{b^2}{2} f_{8AB} f^{8CD})\mathcal{B}_{C}\wedge \mathcal{B}_D\nn\\
&&-\dot{b}^2(\half f_{8AB}f^{8\dot{A}\dot{B}}+\frac{1}{2}f^{3AB}f_{3\dot{A}\dot{B}}) \mathcal{B}^{\dot{A}}\wedge \mathcal{B}^{\dot{B}}\nn\\
&&+\Big(-\frac{b^2}{2c^2}f^{3AB}-\frac{\sqrt{3}b^2}{c^2} f^{8AB}\Big)\nn\\
&&\Big(\frac{a^2}{2} f_{3mn} \mathcal{B}^m\wedge \mathcal{B}^n+(\frac{b^2}{2}f_{3CD} 
+\sqrt{3}b^2f^{8CD})\mathcal{B}_{C}\wedge \mathcal{B}_D+(\frac{\dot{b}^2}{2}f_{3\dot{C}\dot{D}}+\sqrt{3}\dot{b}^2 f^{8\dot{C}\dot{D}})\mathcal{B}_{\dot{C}}\wedge \mathcal{B}_{\dot{D}}\Big)\nn\\
&&-\Big(-\frac{b\dot{b}}{2a}
+\frac{a\dot{b}}{2b}
+\frac{ab}{2\dot{b}}\Big)f^{mA\dot{C}} \Big(-\frac{b\dot{b}}{2a}
+\frac{a\dot{b}}{2b} 
+\frac{ab}{2\dot{b}}\Big)f^{nB\dot{C}}\mathcal{B}^m\wedge\mathcal{B}^n\nn\\
&&-(\frac{b^2}{2c}f^{3AC}+\frac{\sqrt{3}b^2}{c}f^{8AC}) (\frac{b^2}{2c}f^{3BD}+\frac{\sqrt{3}b^2}{c}f^{8BD})\mathcal{B}^{C}\wedge\mathcal{B}^{D}\nn\\
&&-\Big[\frac{b\dot{b}}{2a}
+\frac{a\dot{b}}{2b}
-\frac{ab}{2\dot{b}}\Big]f_{mA\dot{B}} \Big[\frac{b\dot{b}}{2a}
+\frac{a\dot{b}}{2b}
-\frac{ab}{2\dot{b}}\Big] f^{mB\dot{C}}\mathcal{B}^{\dot{B}}\wedge\mathcal{B}^{\dot{C}}
\ee 
From the above expression one can read off the components of the Riemann tensor in \eqref{eqn:riemann_components} given below.
	\be 
&&\mathcal{R}^{AZ}_{CZ}=-(\frac{b^2}{2c}f^{3AB}+\frac{\sqrt{3}b^2}{c}f^{8AB}) (\frac{b^2}{2c}f^{3BC}+\frac{\sqrt{3}b^2}{c}f^{8BC})\mathcal{B}^C\wedge\mathcal{B}^{Z}
=+\frac{49b^4}{16c^2}\nn
\ee

\be 
&&\mathcal{R}^{AB}_{CD}=-\frac{1}{2}(\frac{b^2}{2}f^{3AB}f_{3CD}+\frac{b^2}{2} f_{8AB} f^{8CD})
+\frac{1}{2}\Big(-\frac{b^2}{2c^2}f^{3AB}-\frac{\sqrt{3}b^2}{c^2} f^{8AB}\Big)(\frac{b^2}{2}f_{3CD} 
+\sqrt{3}b^2f^{8CD}) \nn\\
&&-\frac{1}{2}(\frac{b^2}{2c}f^{3AC}+\frac{\sqrt{3}b^2}{c}f^{8AC}) (\frac{b^2}{2c}f^{3BD}+\frac{\sqrt{3}b^2}{c}f^{8BD})\nn
\ee

\be 	\mathcal{R}^{A\dot{B}}_{C\dot{D}}&&=\frac{1}{2}\Big(-\frac{b^2\dot{b}^2}{2a^2}
+\frac{\dot{b}^2}{2}
+\frac{b^2}{2}\Big)\Big(f_{mA\dot{B}}f_{mC\dot{D}} \Big)
+\frac{1}{2}\Big[\frac{b\dot{b}}{2a}
+\frac{a\dot{b}}{2b}
-\frac{ab}{2\dot{b}}\Big]f^{mA\dot{D}} \Big[-\frac{b\dot{b}}{2a}
+\frac{a\dot{b}}{2b}
-\frac{ab}{2\dot{b}}\Big]f^{mC\dot{B}}\nn\\
&&-\frac{1}{2}(\frac{b^2}{2c}f^{3AC}+\frac{\sqrt{3}b^2}{c}f^{8AC})(\frac{\dot{b}^2}{2c}f^{3\dot{B}\dot{D}}
+\frac{\sqrt{3}\dot{b}^2}{c} f^{8\dot{B}\dot{D}})\nn
\ee 

\be 
\mathcal{R}^{Am}_{Bn}&=&-\frac{1}{2}\Big[-\frac{b^2}{2}
-\frac{a^2}{2}
+\frac{a^2b^2}{2\dot{b}^2}\Big]f^{mA\dot{B}}f^{nB\dot{B}}+\frac{1}{2}\frac{a^2}{2c}f^{3nm}(\frac{b^2}{2c}f^{3AB}+\frac{\sqrt{3}b^2}{c}f^{8AB})\nn\\
&&+\frac{1}{2}\Big(-\frac{b\dot{b}}{2a}
+\frac{a\dot{b}}{2b} 
+\frac{ab}{2\dot{b}}\Big)f^{nA\dot{B}}f^{mB\dot{B}} \Big[-\frac{b\dot{b}}{2a}
+\frac{a\dot{b}}{2b} 
-\frac{ab}{2\dot{b}}\Big]\nn
\ee 
Contracting the second and fourth indices of the Riemann tensor we compute the Ricci tensor.
\be 
&&\mathcal{R}^{AB}_{CB}=-\frac{1}{2}(\frac{b^2}{2}f^{3AB}f_{3CB}+\frac{b^2}{2} f_{8AB} f^{8CB})
+\frac{1}{2}\Big(-\frac{b^2}{2c^2}f^{3AB}-\frac{\sqrt{3}b^2}{c^2} f^{8AB}\Big)(\frac{b^2}{2}f_{3CB} 
+\sqrt{3}b^2f^{8CB}) \nn\\
&&-\frac{1}{2}(\frac{b^2}{2c}f^{3AC}+\frac{\sqrt{3}b^2}{c}f^{8AC}) (\frac{b^2}{2c}f^{3BB}+\frac{\sqrt{3}b^2}{c}f^{8BB})
=-\frac{b^2}{4}
-\frac{49b^4}{32c^2} \nn
\ee	
\be 
\mathcal{R}^{A\dot{D}}_{C\dot{D}}&&=
\frac{\dot{b}^2}{8}
+\frac{b^2}{8}
+\frac{a^2\dot{b}^2}{16b^2}
-\frac{a^2}{8}
+\frac{a^2b^2}{16\dot{b}^2}-\frac{3b^2\dot{b}^2}{16a^2}
\ee	
\be 
\mathcal{R}^{Am}_{Bm}&=&-\frac{1}{2}\Big[-\frac{b^2}{2}
-\frac{a^2}{2}
+\frac{a^2b^2}{2\dot{b}^2}\Big]f^{mA\dot{B}}f^{mB\dot{B}}+\frac{1}{2}\frac{a^2}{2c}f^{3mm}(\frac{b^2}{2c}f^{3AB}+\frac{\sqrt{3}b^2}{c}f^{8AB})\nn\\
&&+\frac{1}{2}\Big(-\frac{b\dot{b}}{2a}
+\frac{a\dot{b}}{2b} 
+\frac{ab}{2\dot{b}}\Big)f^{mA\dot{B}}f^{mB\dot{B}} \Big[-\frac{b\dot{b}}{2a}
+\frac{a\dot{b}}{2b} 
-\frac{ab}{2\dot{b}}\Big]\nn\\ [3mm]
&=&\frac{b^2}{8}
+\frac{a^2}{8}
-\frac{a^2b^2}{8\dot{b}^2}
+\frac{b^2\dot{b}^2}{16a^2}
-\frac{\dot{b}^2}{8}
+\frac{a^2\dot{b}^2}{16b^2}
-\frac{a^2b^2}{16\dot{b}^2}
\ee 
Combining all the components above we find $\mathcal{R}^A_C$ below,
\be 
&& \mathcal{R}^A_C={\mathcal{R}^{Am}_{Cm}}+\mathcal{R}^{AZ}_{CZ}+R^{AB}_{CB}+\mathcal{R}^{A\dot{B}}_{C\dot{B}}=
-\frac{a^2b^2}{8\dot{b}^2}
+\frac{a^2\dot{b}^2}{8b^2}
-\frac{b^2\dot{b}^2}{8a^2}\nn 
\ee			
Similarly, the remaining three components of the Ricci tensor are determined and are provided below,
\be 
&&\mathcal{R}^A_C=\Big(-\frac{a^2b^2}{8\dot{b}^2}
	+\frac{a^2\dot{b}^2}{8b^2}
	-\frac{b^2\dot{b}^2}{8a^2}\Big)\delta^A_C\qquad A,C=4,5\nn\\
&&\mathcal{R}^{\dot{A}}_{\dot{C}}=\Big(\frac{a^2b^2}{8\dot{b}^2}-\frac{a^2\dot{b}^2}{8b^2}
	-\frac{b^2\dot{b}^2}{8a^2}\Big) \delta^{\dot{A}}_{\dot{C}}\qquad \dot{A},\dot{C}=6,7\nn\\
&&\mathcal{R}^m_n=\Big[
	\frac{b^2\dot{b}^2}{8a^2}
	-\frac{a^2b^2}{8\dot{b}^2}
	-\frac{a^2\dot{b}^2}{8b^2}\Big]\delta^m_n\qquad m,n=1,2\nn\\
&&\mathcal{R}^Z_Z=\frac{a^4}{4c^2}+\frac{49b^4}{16c^2}+\frac{25\dot{b}^4}{16c^2}
\ee 
Out of many solutions to the above equations, we choose the following solution to the above equations,
\be 
a =4\sqrt{6},\quad b=4\sqrt{6},\quad c=12\sqrt{26},\quad \dot{b}=-4\sqrt{6}\;.
\ee 
%\section{Young tableau}

	%%%%%%%%%%%%%
	
	%%%%%%%%%%%%%%%%%%%%%%%%%%%

\end{document}